\documentclass[prd,preprint,tightenlines,showpacs,preprintnumbers,nofootinbib,eqsecnum,superscriptaddress]{revtex4-1}

 \usepackage[dvips,final]{graphicx}
  \usepackage{amssymb}
   \usepackage{amsmath}
    \usepackage{amsfonts}
     \usepackage{epsfig}
      \usepackage{bm}% bold math

\usepackage[section]{placeins}

\usepackage{multirow}
\usepackage{ctable}
\usepackage{booktabs}
\usepackage{array}
\usepackage{tabularx}
\usepackage{xcolor}
\usepackage{pstricks}
\newcommand{\bp}{\mbox{\boldmath $p$}}
\newcommand{\bq}{\mbox{\boldmath $q$}}

\newcommand{\bk}{\mbox{\boldmath $k$}}

\newcommand{\Tr}{{\mbox{\rm Tr}}}
\begin{document}

%\vfill

\title{Prompt hadroproduction of $\eta_c(1S,2S)$ in the $k_T$-factorization approach}

\author{Izabela Babiarz}
\email{izabela.babiarz@ifj.edu.pl}
\affiliation{Institute of Nuclear Physics, Polish Academy of Sciences, 
ul. Radzikowskiego 152, PL-31-342 Krak{\'o}w, Poland}

\author{Roman Pasechnik}
\email{roman.pasechnik@thep.lu.se}
\affiliation{Department of Astronomy and Theoretical Physics,
Lund University, SE-223 62 Lund, Sweden}

\author{Wolfgang Sch\"afer}
\email{wolfgang.schafer@ifj.edu.pl} 
\affiliation{Institute of Nuclear
Physics, Polish Academy of Sciences, ul. Radzikowskiego 152, PL-31-342 
Krak{\'o}w, Poland}

\author{Antoni Szczurek}
\email{antoni.szczurek@ifj.edu.pl}
\affiliation{College
of Natural Sciences, Institute of Physics,
University of Rzesz\'ow, ul. Pigonia 1, PL-35-310 Rzesz\'ow, Poland\vspace{1cm}}
%\footnote{also at Institute of Nuclear Physics, Kraków, Poland}}

\begin{abstract}
\vspace{0.5cm}
In this work, we present a thorough analysis of $\eta_c(1S,2S)$ 
quarkonia hadroproduction in $k_\perp$-factorisation in the framework of the light-front potential approach for the quarkonium wave function. 
The off-shell matrix elements for the $g^* g^* \eta_c(1S,2S)$ vertices
are derived. We discuss the importance of taking into account the gluon virtualities.
%We discuss the role of relativistic corrections, the Melosh spin-rotation and theoretical uncertainties by comparing our results with the standard NRQCD approach. 
We present the transverse momentum distributions of $\eta_c$ for several models of the unintegrated gluon distributions. Our calculations are performed for four distinct parameterisations for the $c\bar c$ interaction potential consistent with the meson spectra.
We compare our results for $\eta_c(1S)$ to measurements by the
LHCb collaboration and present predictions for $\eta_c(2S)$ production.
\end{abstract}

\pacs{12.38.Bx, 13.85.Ni, 14.40.Pq}
\maketitle

%----------------------------
\section{Introduction}
%----------------------------

The quarkonia production reactions in hadronic collisions at the Large Hadron Collider (LHC) continue to attract a lot of interest \cite{Lansberg:2019adr}. In this paper, we concentrate on the direct
hadroproduction of the ground state of the charmonium family, $\eta_c(1S)$, and its first excited state $\eta_c(2S)$. Both are pseudoscalar particles of even charge parity $J^{PC} = 0^{-+}$. 
Like other $C$-even quarkonia, the dominant production mechanism is through the $gg \to {\cal{Q}}$ gluon fusion $2 \to 1$ process.
In the standard collinear-factorization approach one must go to next-to-leading order (NLO) approximation to calculate the transverse momentum distribution of a given quarkonium state and include $ 2 \to 2$ processes like $g g \to {\cal{Q}} g$. In the $k_T$-factorization approach \cite{UGD_GLR,UGD_CCH,UGD_CE}, the transverse momentum of the quarkonium originates from the transverse momenta of incident virtual gluons entering the hard $g^* g^* \to {\cal{Q}}$ process. 
The $k_T$-factorization approach is especially appropriate in the high-energy kinematics, where partons carry small momentum fractions $x$ of the incoming protons, often discussed in the framework of the BFKL formalism \cite{BFKL}. In our calculations we will adopt the color-singlet model, which treats the quarkonium as a two-body bound state of a heavy quark and antiquark. Such a formalism was used previously for the production of $\chi_{cJ}$ ($J=0,1,2$) quarkonia (see e.g. Ref.~\cite{CS2018}), and a relatively good agreement with data was obtained from an unintegrated gluon distribution (UGD), which effectively includes the higher-order contributions.

Recently, the LHCb collaboration has measured the transverse momentum
distributions of $\eta_c$ in the $p \bar p$ decay channel
\cite{Aaij:2014bga} (see also the recent PhD thesis \cite{Usachov:2019czc}). The experimental method allows to measure $\eta_c$ charmonia only for $p_T > 6.5 \, \rm{GeV}$. In the present study, we will discuss production of $\eta_c(1S)$ and $\eta_c(2S)$ also at lower transverse momenta. This is a region where the effects of nonlinear evolution for the UGDs may potentially show up. This was discussed briefly in Ref.~\cite{CS2018} in the context of low transverse momentum $\chi_c$ production.

A crucial ingredient of our $k_T$-factorization approach is the off-shell matrix element for the $g^* g^* \to \eta_c$ transition. Recently in Ref.~\cite{Babiarz:2019sfa} we discussed in detail the $\gamma^* \gamma^* \to \eta_c(1S,2S)$ form factors. These form factors were calculated there from the $c \bar c$ light-front (LF) wave functions obtained from different $c \bar c$ interaction potentials obtained in Ref.~\cite{Cepila:2019skb}. In the present paper, we will apply the formalism developed in Ref.~\cite{Babiarz:2019sfa} using the potential approach and the Melosh spin transform to derive the proper LF wave functions of the $\eta_c(1S,2S)$ states\footnote{For a recent analysis of the role of the Melosh spin transform in vector $S$-wave meson photoproduction with and without $D$-wave admixture, see Ref.~\cite{Krelina:2019egg}}. Here, we employ the same LF formalism for computation of the $g^* g^* \to \eta_c$ vertex, for both gluons being off-shell. We wish to demonstrate the role of the form factor and to estimate the uncertainties for the $\eta_c$ production yields related to it. This was not discussed so far in the context of quarkonia production in proton-proton collisions. We wish to focus also on a possibility of testing the unintegrated gluon distributions by comparing our predictions to the experimental data.

Previously, the prompt $\eta_c(1S)$ production was discussed in various factorization approaches: collinear factorization 
\cite{Diakonov:2012vb,Likhoded:2014fta,Butenschoen:2014dra,Zhang:2014ybe,Feng:2019zmn}, the TMD-factorization with transverse momentum dependent distributions of on-shell gluons, and on-shell matrix elements \cite{Boer:2012bt,Echevarria:2019ynx}, and the $k_T$-factorization in Ref.~\cite{Baranov:2019joi}. So far, the $\eta_c(2S)$ production process was discussed only in the collinear factorization approach in Ref.~\cite{Lansberg:2017ozx}.

The paper is organised as follows. In Section~\ref{Sect:Formalism}, we discuss the formalism behind the quarkonia hadroproduction processes in the $k_T$ factorization approach. In Section~\ref{Sect:numerics} we present the most relevant numerical results for the differential $\eta_c(1S,2S)$ production cross sections versus the available experimental data and discuss the related theoretical uncertainties. The basic concluding remarks and the main results are summarised in Section~\ref{Sect:conclusions}.
 
%----------------------------
\section{Formalism}
\label{Sect:Formalism}
%----------------------------

%----------------------------------------------------------------------------
\subsection{Off-shell matrix element and cross section}
%----------------------------------------------------------------------------

\begin{figure}
    \centering
    \includegraphics[width=0.6\textwidth]{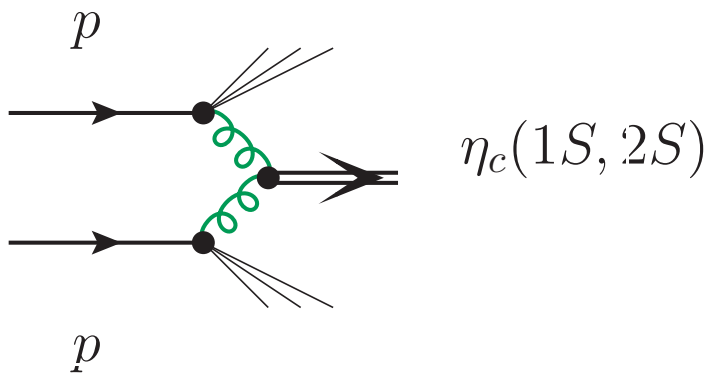}
    \caption{Generic diagram for the inclusive process of 
$\eta_{c}$(1S) or $\eta_{c}$(2S) production in proton-proton scattering 
via two gluons fusion. }
    \label{fig:diagram_etac}
\end{figure}

In Fig.\ref{fig:diagram_etac} we show a generic Feynman diagram for $\eta_c(1S)$ quarkonium production in proton-proton collision via gluon-gluon fusion.
This diagram illustrates the situation adequate for the
$k_T$-factorization calculations used in the present paper.
%--------------------------------------------------------------------------
%---------------------------------------------------------------------------
The inclusive cross section for $\eta_c$-production via the $2 \to 1$ gluon-gluon fusion mode is obtained from
%%%
\begin{eqnarray}
d\sigma = \int {dx_1 \over x_1} \int {d^2 \bq_1 \over \pi \bq_1^2} 
{\cal{F}}(x_1,\bq_1^2,\mu_F^2)\int {dx_2 \over x_2} 
\int {d^2 \bq_2 \over \pi \bq_2^2}  {\cal{F}}(x_2,\bq_2^2,\mu_F^2) {1 \over 2 x_1 x_2 s} \overline{|{\cal{M}}|^2} \, d\Phi(2 \to 1). \nonumber \\
\end{eqnarray}
%%%%
The unintegrated gluon distributions are normalized such, that the collinear glue is obtained from
%%%%
\begin{eqnarray}
xg(x,\mu_F^2) = \int^{\mu_F^2} {d \bk^2 \over \bk^2} {\cal{F}}(x,\bk^2,\mu_F^2) \, ,
\end{eqnarray}
%%%%
where from now on we will no longer show the dependence on the factorization scale $\mu_F^2$ explicitly.
%\begin{eqnarray}
%{\cal{F}}(x,\bq^2) = {\partial x g(x,\bq^2) \over %\partial \log \bq^2}
%\end{eqnarray}
%%%%
Let us denote the four-momentum of the $\eta_c$ by $p$
and parametrize it in light-cone coordinated as
%%%
\begin{eqnarray}
p = (p_+,p_-,\bp) = ({m_T \over \sqrt{2}} e^y, {m_T \over \sqrt{2}} e^{-y}, \bp) \, , 
\end{eqnarray}
%%%
where we introduced the transverse mass
%%%
\begin{eqnarray}
m_T = \sqrt{\bp^2 + m^2} ,
\end{eqnarray}
%%%
where $m$ is the mass of the $\eta_c$-meson, and $y$ is its rapidity in the $pp$ cms-frame.
The phase-space element is
%%%%
\begin{eqnarray}
d\Phi (2 \to 1) = (2 \pi)^4 \delta^{(4)}(q_1 + q_2 - p) \, {d^4 p \over (2 \pi)^3} \delta(p^2 - m^2) \, .
\end{eqnarray}
%%%%
In the $k_T$-factorization approach, gluons are off-shell,
$q_i^2 = - \bq_i^2$, and their four momenta are written as
($\sqrt{s}$ is the pp center-of-mass energy):
 %%%
\begin{eqnarray}
q_1 = (q_{1+},0,\bq_1) \, , \, q_2 = (0,q_{2-},
 \bq_2) \, ,
\end{eqnarray}
%%%
with 
\begin{eqnarray}
q_{1+} = x_1 \sqrt{s \over 2} \, , \, q_{2-} = x_2  \sqrt{s \over 2} \, .
\end{eqnarray}
%%%%%
We can then calculate the phase-space element as
%%%%% 

%\begin{eqnarray}
%d\Phi (2 \to 1) = 2 \pi \delta(q_{1_+} - p_+) \delta(q_{2-} - p_-) \delta^{(2)}(\bq_1 + \bq_2 - \bp) \, dp_+ dp_- d^2\bp \, \delta( 2 p_+ p_- - \bp^2 - m^2) \, . \nonumber \\
%\end{eqnarray}
%%%%
%This gives
%%%%%
\begin{eqnarray}
d\Phi(2 \to 1) &=& 
%2 \pi \, {2 \over s} \, \delta(x_1 - {m_\perp \over \sqrt{s}} e^y)\delta(x_2 - {m_\perp \over \sqrt{s}} e^{-y}) \delta^{(2)}(\bq_1 + \bq_2 - \bP) {dP_+ \over 2P_+} d^2\bP \nonumber \\
{2 \pi \over s} \delta(x_1 - {m_T \over \sqrt{s}} e^y)\delta(x_2 - {m_T \over \sqrt{s}} e^{-y}) \delta^{(2)}(\bq_1 + \bq_2 - \bp) \,  dy \, d^2\bp \, .
\end{eqnarray}
%------------------------------------------------------------------------
We therefore obtain for the inclusive cross section
%%%%
\begin{eqnarray}
{d \sigma \over dy d^2\bp} = \int {d^2 \bq_1 \over \pi \bq_1^2} 
{\cal{F}}(x_1,\bq_1^2) \int {d^2 \bq_2 \over \pi \bq_2^2}  {\cal{F}}(x_2,\bq_2^2) \, \delta^{(2)} (\bq_1 + \bq_2 - \bp ) \, {\pi \over (x_1 x_2 s)^2} \overline{|{\cal{M}}|^2} \,  ,
\end{eqnarray}
%%%%
where the momentum fractions of gluons are fixed as $x_{1,2} = m_T \exp(\pm y) / \sqrt{s}$.
The off-shell matrix element is written in terms of the Feynman amplitude
as (we restore the color-indices):
%%%%
\begin{eqnarray}
{\cal{M}}^{ab} = {q_{1 \perp}^\mu q_{2\perp}^\nu \over |\bq_1| |\bq_2|}{\cal{M}}^{ab}_{\mu \nu}  = {q_{1+} q_{2-} \over |\bq_1| |\bq_2|} n^+_\mu n^-_\nu {\cal{M}}^{ab}_{\mu \nu} = {x_1 x_2 s \over 2 |\bq_1| |\bq_2| } n^+_\mu n^-_\nu {\cal{M}}^{ab}_{\mu \nu}  \, .
\end{eqnarray}
%%%%
In covariant form, the matrix element reads: %%%
\begin{eqnarray}
{\cal{M}}^{ab}_{\mu \nu} = (-i) 4 \pi \alpha_s \, \varepsilon_{\mu \nu \alpha \beta} q_1^\alpha q_2^\beta {\Tr[t^a t^b] \over \sqrt{N_c}} \, I(\bq_1^2,\bq_2^2) \, .
\end{eqnarray}
%%%
To the lowest order, it is proportional to the matrix element for the $\gamma^* \gamma^* \eta_c$ vertex.
In particular, the form factor $I(\bq_1^2,\bq_2^2)$ is related to the $\gamma^* \gamma^* \eta_c$ transition form factor $F(Q_1^2,Q_2^2), \, Q_i^2 = \bq_i^2$ as
%%%
\begin{eqnarray}
F(Q_1^2,Q_2^2) = e_c^2 \sqrt{N_c} \, I(\bq_1^2,\bq_2^2) \, ,
\end{eqnarray}
%%%%
and it can be represented in terms of the LF wave function as
%%%
\begin{eqnarray}
I(\bq_1^2,\bq_2^2) &=&   4 m_c 
\int {dz d^2 \bk \over z(1-z) 16 \pi^3} \psi(z,\bk) 
\Big\{ 
{1-z \over (\bk - (1-z) \bq_2 )^2  + z (1-z) \bq_1^2 + m_c^2}
\nonumber \\
&+& {z \over (\bk + z \bq_2 )^2 + z (1-z) \bq_1^2 + m_c^2}
\Big\} \, .
\label{eq:FF}
\end{eqnarray}
%%%%
For details of the derivation and the normalization conventions and relation to the potential model wave function of the LF radial wave function $\psi(z,\bk)$, see Ref.~\cite{Babiarz:2019sfa}. 
%%%%
 In this work, we will use the calculations of the form factor which were obtained in \cite{Babiarz:2019sfa}. There, the representation of the $\gamma^* \gamma^* \eta_c$ transition form factor in terms
 of the LF wave function of the $\eta_c$ was derived. Several wave-functions
 obtained from potential models for the $c \bar c$ system which were previously obtained in Ref.~\cite{Cepila:2019skb} were used.
 
%Then, we obtain for the cross section
%%%%
%\begin{eqnarray}
%{d \sigma \over dy d^2\bP} = \int {d^2 \bq_1 \over \pi \bq_1^4} 
%{\cal{F}}(x_1,\bq_1^2) \int {d^2 \bq_2 \over \pi \bq_2^4}  {\cal{F}}(x_2,\bq_2^2) \, \delta^{(2)} (\bq_1 + \bq_2 - \bP ) \, {\pi \over 4} \overline{|n^+_\mu n^-_\mu{\cal{M}}_{\mu \nu}|}^2 ,
%\end{eqnarray}
%%%%
%The matrix element squared averaged over color is
%%%%
%\begin{eqnarray}
%\overline{|n^+_\mu n^-_\mu{\cal{M}}_{\mu \nu}|}^2 = {1 \over (N_c^2 -1)^2}
%\sum_{a,b} |n^+_\mu %n^-_\mu{\cal{M}}^{ab}_{\mu \nu}|^2
%\end{eqnarray}
%%%
Inserting the explicit form of the matrix element
%%%
\begin{eqnarray}
n^+_\mu n^-_\mu{\cal{M}}^{ab}_{\mu \nu} &=& 4 \pi \alpha_s  (-i) [\bq_1,\bq_2]
{\Tr[t^a t^b] \over \sqrt{N_c}} \, I(\bq_1^2,\bq_2^2) \, 
%\nonumber \\
= 4 \pi \alpha_s (-i) {1 \over 2} \delta^{ab} {1 \over \sqrt{N_c}}  [\bq_1,\bq_2] \, I(\bq_1^2,\bq_2^2) \, ,
\nonumber \\
\end{eqnarray}
%%%
%It is related to the $\gamma^* \gamma^* \eta_c$ transition form factor 
%through the relation
%%%
%\begin{eqnarray}
%F(Q_1^2,Q_2^2) = e_c^2 \sqrt{N_c} \, I(\bq_1^2,\bq_2^2) \,.
%\end{eqnarray}
%%%%
and averaging over colors, we obtain our final result:
%%%%%
\begin{eqnarray}
{d \sigma \over dy d^2\bp} = \int {d^2 \bq_1 \over \pi \bq_1^4} 
{\cal{F}}(x_1,\bq_1^2) \int {d^2 \bq_2 \over \pi \bq_2^4}  {\cal{F}}(x_2,\bq_2^2) \, \delta^{(2)} (\bq_1 + \bq_2 - \bp ) \, {\pi^3 \alpha_s^2 \over N_c (N_c^2-1)}  |[\bq_1,\bq_2] \, I(\bq_1^2,\bq_2^2)|^2  .
\nonumber \\
\end{eqnarray}
Parametrizing the transverse momenta as $\bq_i = (q_i^x, q_i^y) = |\bq_i|(\cos \phi_i, \sin \phi_i)$, we can write the vector product $[\bq_1,\bq_2]$ as
%%%%
\begin{eqnarray}
[\bq_1,\bq_2] = q_1^x q_2^y - q_1^y q_2^x = |\bq_1| |\bq_2| 
\sin(\phi_1 - \phi_2) \, .
\end{eqnarray}
%%%%
%Then, the averaged matrix element squared becomes
%%%%
%\begin{eqnarray}
%\overline{|n^+_\mu n^-_\mu{\cal{M}}_{\mu \nu}|}^2
%&=& 16 \pi^2 \alpha_s^2 {1 \over 4} {1 \over N_c} |[\bq_1,\bq_2] \, I(\bq_1^2,\bq_2^2)|^2   {1 \over (N_c^2 -1)^2} \sum_{a,b} \delta^{ab} \delta^{ab} \nonumber \\
%&=& 4 \pi^2 \alpha_s^2 {1 \over N_c (N_c^2 -1)}  |[\bq_1,\bq_2] \, I(\bq_1^2,\bq_2^2)|^2 
%\end{eqnarray}
%%%%%
In our numerical calculations presented below, we set the factorization scale to $\mu_F^2 = m_T^2$, and the renormalization scale is taken in the form:
\begin{equation}
\alpha_s^2 \to 
\alpha_s(\max{\{m_T^2,\bq_1^2\}})
\alpha_s(\max{\{m_T^2,\bq_2^2\}})  \; .
\label{alpha_s}
\end{equation}
%
%%%%
%------------------------------------------------------------------------------
\subsection{Normalization of the $g^* g^* \eta_c(1S,2S)$ form factors}
%------------------------------------------------------------------------------
%%%
The normalization of the inclusive cross section depends crucially on the value of the
$g^* g^* \eta_c$ form factor for vanishing gluon virtualities $\bq_1^2 = \bq_2^2 = 0$.
The latter in turn is directly related to the $\eta_c \to gg$ decay width.
From the proportionality of the $g^* g^* \eta_c$ and $\gamma^* \gamma^* \eta_c$ vertices to the leading order (LO), we obtain, that at LO, the $\gamma \gamma$ and $gg$ widths are related by
%%%%%
\begin{eqnarray}
\label{eq:LO_gluon}
\Gamma_{\rm{LO}}(\eta_c \to gg) = {N_c^2 -1 \over 4 N_c^2} \, {1 \over e_c^4} \,
\Big({\alpha_s \over \alpha_{\rm em}} \Big)^2 \, \Gamma_{\rm{LO}}(\eta_c \to \gamma \gamma) \, ,
\end{eqnarray}
%%%
where the LO $\gamma \gamma$ width in turn is related to the
transition form factor for vanishing virtualities through
%%%%
\begin{equation}
\label{eq:L0_photon}
    \Gamma_{\rm{LO}}(\eta_c \to \gamma \gamma) = {\pi \over 4} \alpha^2_{\rm em}
    M^3_{\eta_c} |F(0,0)|^2 \, .
\end{equation}
%%%
At NLO, the expressions for the widths read (see e.g. \cite{Lansberg:2006dw})
%%%%
\begin{eqnarray}
\label{eq:NLO}
\Gamma(\eta_c \to \gamma \gamma) &=& \Gamma_{\rm{LO}}(\eta_c \to \gamma \gamma) \, \Big( 1 - {20 - \pi^2 \over 3} {\alpha_s \over \pi} \Big)\,, \nonumber \\
\Gamma(\eta_c \to g g) &=& \Gamma_{\rm{LO}}(\eta_c \to g g) \, 
\Big( 1 + 4.8 \, {\alpha_s \over \pi} \Big). 
%\nonumber \\
\end{eqnarray}
%%%%%
In order to control the model uncertainty on the normalization, 
one may want to adjust its value $F(0,0)$ to the measured 
decay width. Here we face the ambiguity of fitting either to the hadronic or to the $\gamma \gamma$ width.
As there are no other known radiative decays besides $\gamma \gamma$, 
one may try to identify the $gg$-width with the total (hadronic) width.

In Tables \ref{table:width_tot},\ref{table:width_gamma}, we show the values of $|F(0,0)|$
obtained in three different ways.
In Table~\ref{table:width_tot} we show the result extracted from the total decay width.
Here we use the strong coupling $\alpha_s = 0.26$, which is appropriate to our choice of the renormalization scale in the production amplitudes.
In Table~\ref{table:width_gamma} we extract the value of $|F(0,0)|$ from the
radiative decay width in two different ways.
The first result is obtained based on Eq.(\ref{eq:L0_photon}) 
using the experimental value for $\Gamma(\eta_c \to \gamma \gamma)$ on 
the left hand side, while the second one uses the NLO relation \ref{eq:NLO}.
%By $|F(0,0)|_{gg}$ we show the values extracted from the total width, and by
%$|F(0,0)|_{\gamma \gamma}$ the ones extracted from the $\gamma \gamma$
%width with next-to-leading order precision.

We observe a substantial 
difference between the two different extractions of $|F(0,0)|$. 
While in the $\eta_c(2S)$ case, the error bars are too large to claim an inconsistency, the situation for the $\eta_c(1S)$ is not satisfactory.
This is in fact an old problem and may hint at 
an insufficiency of the potential model treatment of the $\eta_c$. Various possible
solutions have been proposed, such as an admixture of light hadron states \cite{Shifman:1978zq},
a mixing with a pseudoscalar glueball \cite{Kochelev:2005tu}, or nonperturbative instanton effects in the
hadronic decay \cite{Zetocha:2002as}.
%%%%%
\begin{table}[]
    \centering
    \caption{Total decay widths as well as $|F(0,0)|$ obtained from $\Gamma_{tot}$ using the next-to-leading order approximation (see Eq.~(\ref{eq:NLO})).}
    %\label{tab:my_label}
    \begin{tabular}{c|c|c}
    
    \hline
    \hline
     &Experimental values & Derived from Eq.(\ref{eq:NLO}) \\
      
      & $\Gamma_{tot}$ (MeV) \cite{PDG} &$|F(0,0)|_{gg} [GeV^{-1}]$\\ 
                      \hline
    $\eta_{c}(1S)$     & 31.9$\pm$0.7 &0.119$\pm$0.001\\
   $\eta_{c}(2S)$     & 11.3$\pm$3.2$\pm$2.9 &0.053$\pm$0.010 \\
    \hline
    \hline
    \end{tabular}
    \label{table:width_tot}
\end{table}
%%%%%
%%%%%
\begin{table}[]
    \centering
    \caption{Radiative decay widths as well as $|F(0,0)|$ obtained from $\Gamma_{\gamma \gamma}$ using leading order and next-to-leading order approximation (see formulas  Eq. (\ref{eq:L0_photon}, \ref{eq:NLO})).}
    %\label{tab:my_label}
    \begin{tabular}{c|c|c|c}
    \hline
    \hline
      &Experimental values& Derived from Eq.(\ref{eq:L0_photon})&Derived from Eq.(\ref{eq:NLO}) \\
      &  $\Gamma_{\gamma \gamma}$(keV) \cite{PDG} &$|F(0,0)| [GeV^{-1}]$  & $|F(0,0)|_{\gamma \gamma}[GeV^{-1}]$\\ 
                       \hline
    $\eta_{c}(1S)$  & 5.0 $\pm$0.4  & 0.067$\pm$0.003 & 0.079$\pm$0.003 \\
    $\eta_{c}(2S)$  & 1.9 $\pm$1.3 $\cdot$10$^{-4}\cdot \Gamma_{\eta_{c}(2S)}$ & 0.033$\pm$0.012 & 0.038$\pm$0.014\\
    \hline
    \hline
    \end{tabular}
    \label{table:width_gamma}
\end{table}
%%%%%

%-------------------------------------------------------------
\subsection{Unintegrated gluon distributions}
%-------------------------------------------------------------
We use a few different UGDs which are available from the literature, e.g. from the TMDLib package \cite{Hautmann:2014kza} or the CASCADE Monte Carlo code \cite{Jung:2010si}.
\begin{enumerate}
    \item Firstly we use a glue constructed according to the prescription initiated in \cite{Kimber:2001sc} and later updated in \cite{Martin:2009ii}, which we label below as ``KMR''. It uses as an input the collinear gluon distribution from \cite{Harland-Lang:2014zoa}. 
    \item Secondly, we employ two UGDs obtained by Kutak in \cite{Kutak:2014wga}. There are two versions of this UGD. Both introduce a hard scale dependence via a Sudakov form factor into solutions of a small-$x$ evolution equation.
    The first version uses the solution of a linear, BFKL \cite{BFKL} evolution with a resummation of subleading terms and is denoted by ''Kutak (linear)''. The second UGD, denoted as ``Kutak (nonlinear)'' uses instead a nonlinear evolution equation of Balitsky-Kovchegov \cite{BK} type. Both of the Kutak's UGDs can be applied only in the small-$x$ regime, $x < 0.01$.
    \item The third type of UGD which we use has been obtained by Hautmann and Jung \cite{Hautmann:2013tba} from a description of precise HERA data on deep inelastic structure function by a solution of the CCFM evolution equations \cite{CCFM}. We use ``Set 2'' of Ref.~\cite{Hautmann:2013tba}.
\end{enumerate}
%%%%
For the case of the KMR UGD, it has recently been shown in \cite{MS2019}, that it includes 
effectively also higher order corrections of the collinear factorization approach.
In this sense
should give within our approach a result similar to that found recently 
in the NLO approach \cite{Feng:2019zmn} at not too small transverse
momenta. In contrast to the collinear NLO
approach in our approach we can go to very small transverse momenta close to $p_T$ = 0.

%-----------------------------------
\section{Numerical results}
\label{Sect:numerics}
%-----------------------------------

\begin{figure}
    \centering
    \includegraphics[width=0.45\textwidth]{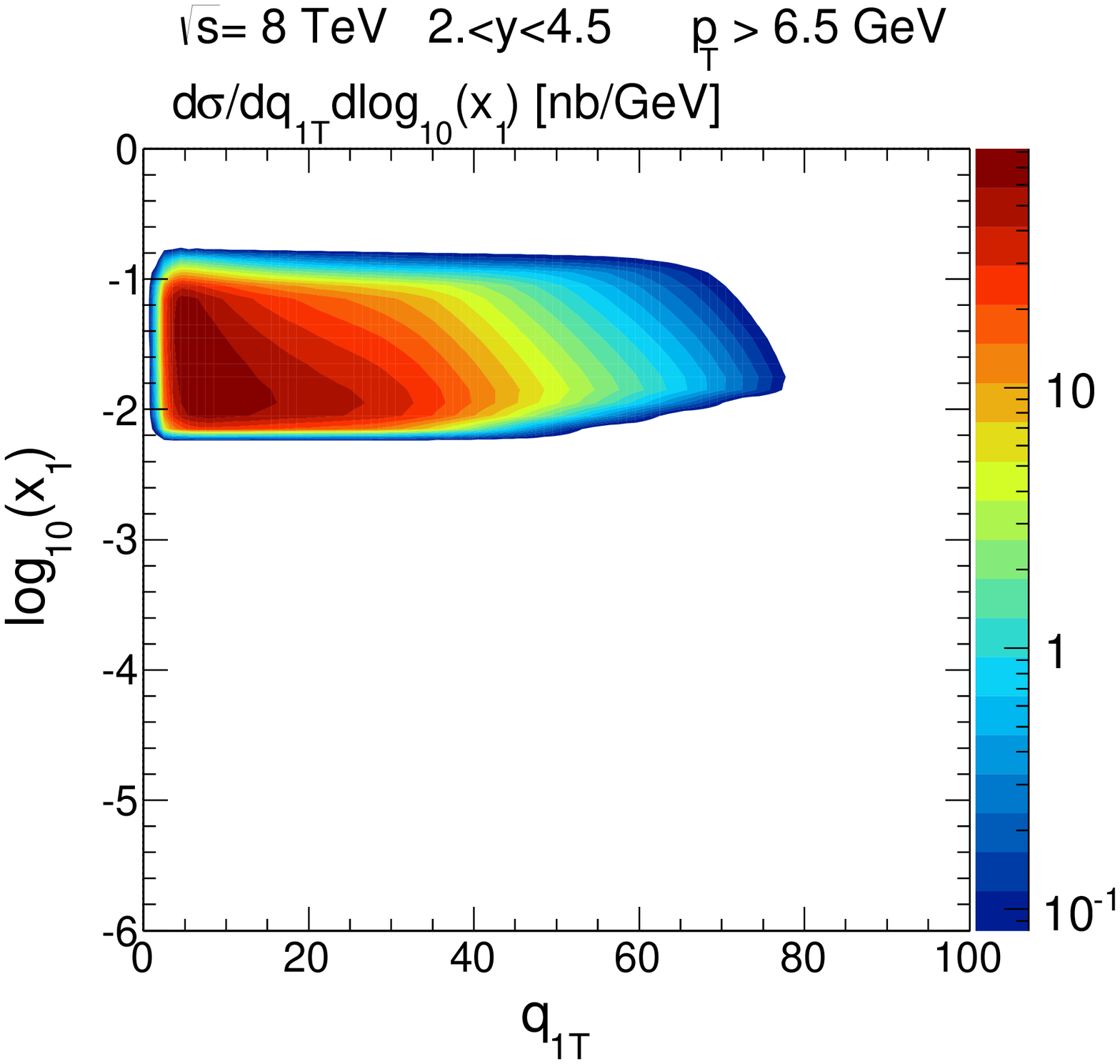}
    \includegraphics[width=0.45\textwidth]{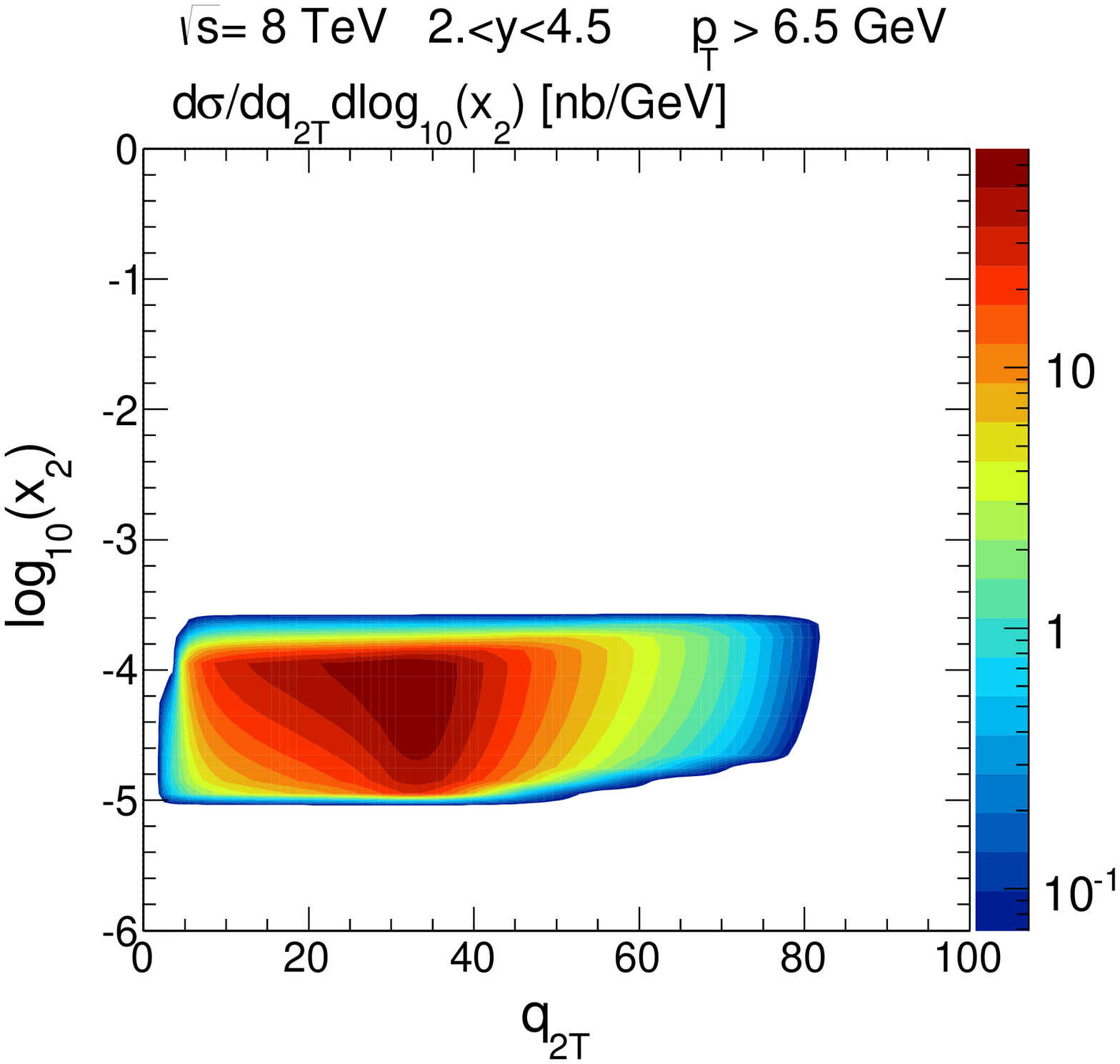}
    \caption{Two-dimensional distributions in $(x_1, q_{1T})$
(left panel) and in $(x_2, q_{2T})$ (right panel) for $\eta_c(1S)$
production for $\sqrt{s}$ = 8 TeV.
In this calculation the KMR UGD was used for illustration.
} 
    \label{fig:dsig_dxdkt}
\end{figure}

Before presenting results for the cross sections let us understand
first the kinematical situation relevant for the LHCb experiment. 
The gluons entering the $g^* g^* \to \chi_c$ vertex 
(see Fig.\ref{fig:diagram_etac})
are characterized by their longitudinal fractions ($x_1$ or $x_2$) 
or gluon virtualities at high-energies directly related
to the their transverse momenta $\bq_{1,2}$.
In Fig.\ref{fig:dsig_dxdkt} we show two dimensional distributions
for the first gluon $(x_1, q_{1T}= |\bq_1|)$ (left panel) and for the second gluon
$(x_2, q_{2T}= |\bq_2|)$ (right panel). We observe a large asymmetry of the two
distributions related to asymmetric LHCb configuration: $y \in (2, 4.5)$. The relatively large lower cut on $\eta_c$ transverse
momentum $p_T > 6.5 \, \rm{GeV}$ causes that $q_{1T}$ and $q_{2T}$ are themselves not
small and are essentially in the perturbative regime where UGDs should be rather reliable.
Moreover, large $p_T$ of the $\eta_c$ also 
entails the large factorization scale.
In this calculation we used the KMR UGD described above. 

%----------------------------------------------------------------------------
%-----------------------------------------------------------------------------

\begin{figure}
    \centering
    \includegraphics[width=0.45\textwidth]{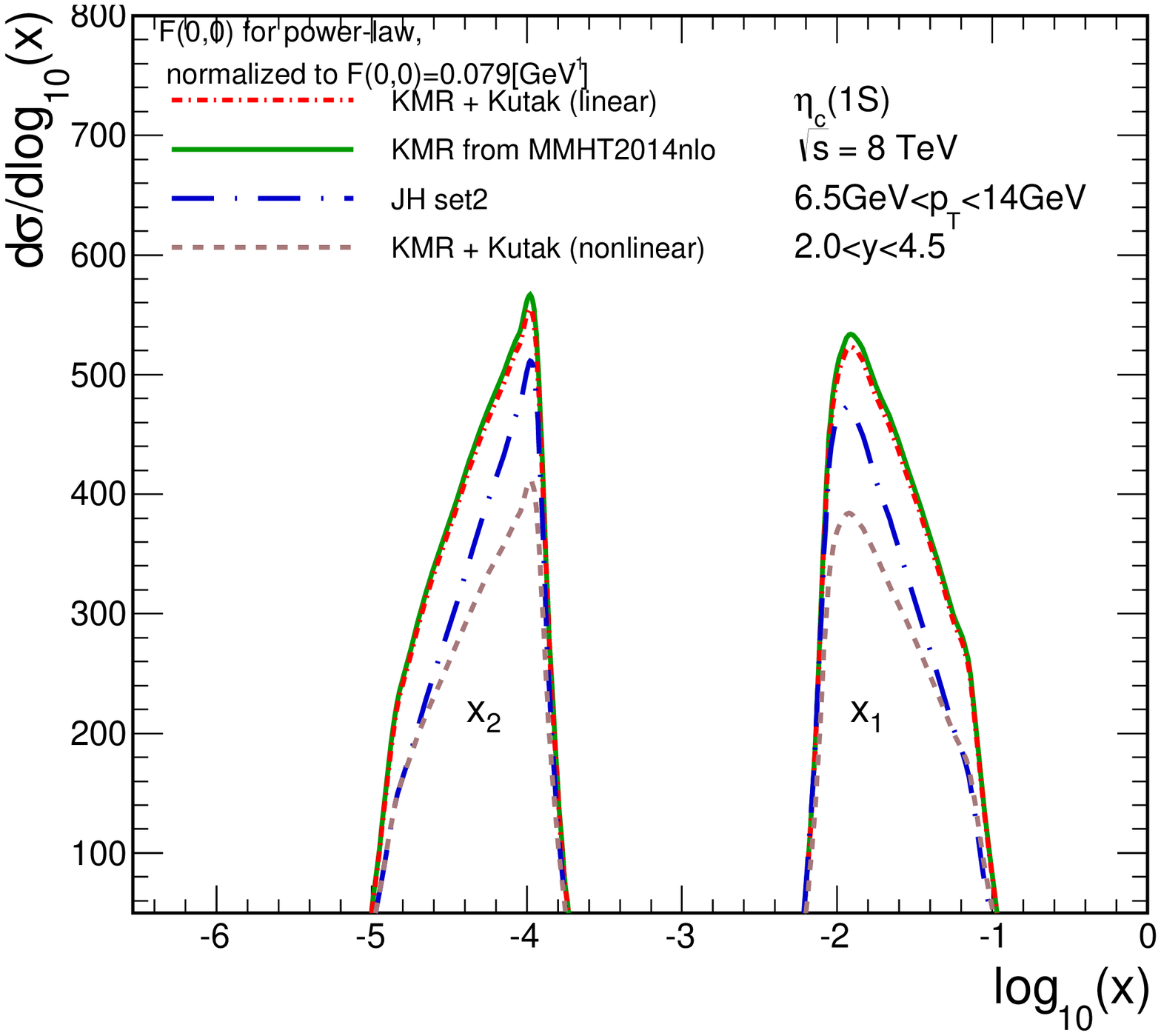}
    \includegraphics[width=0.45\textwidth]{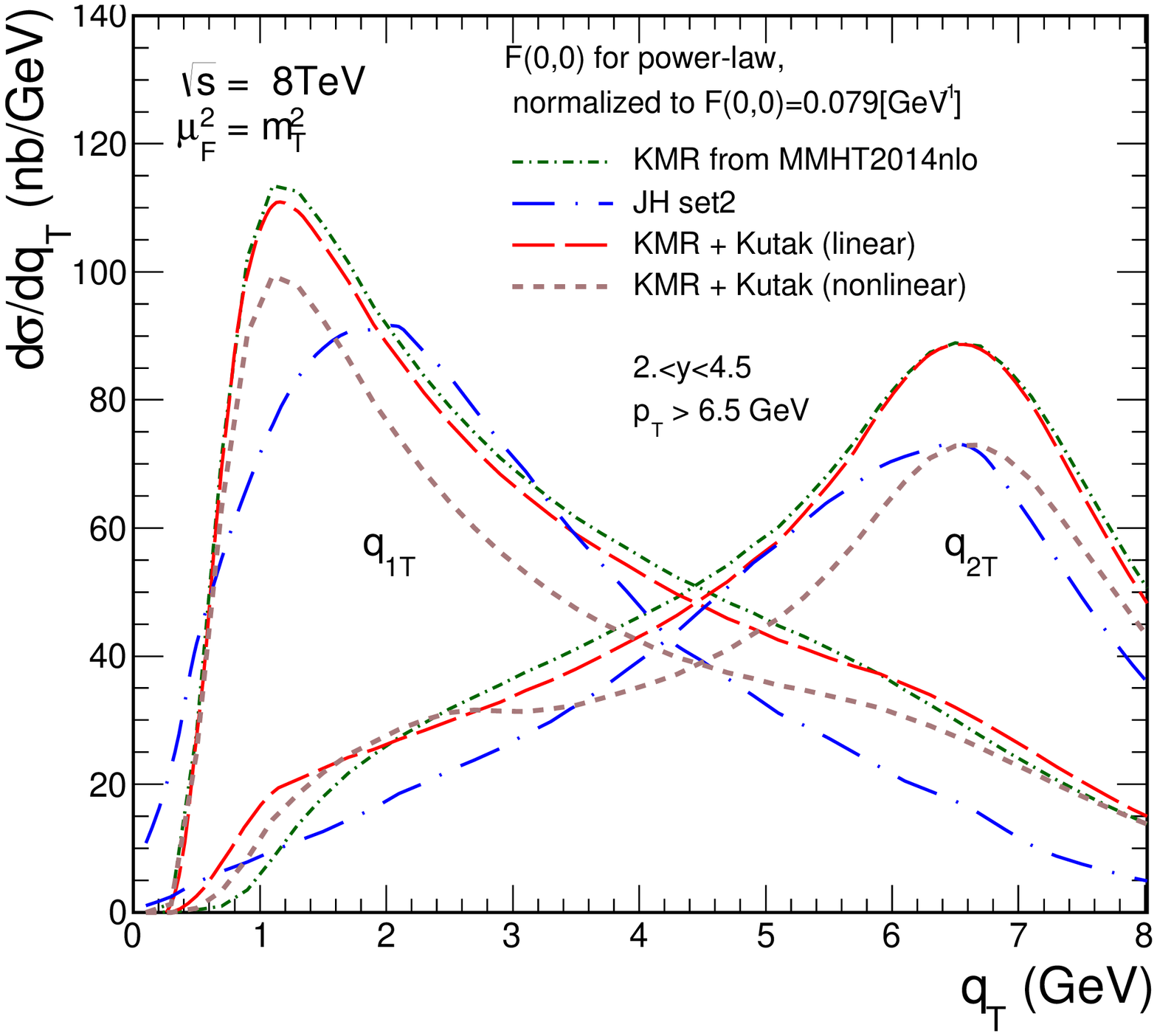}
    \caption{Distributions in  $\log_{10}(x_1)$ or $\log_{10}(x_2)$ 
(left panel) and distributions in $q_{1T}$ or $q_{2T}$    
(right panel) for the LHCb kinematics. Here the different UGDs were used
in our calculations. Here we show an example, where $\sqrt{s}$ = 7 TeV. }
    \label{fig:projections}
\end{figure}

The projections on the $x_i$ and $q_{iT}$ axes are shown in Fig.\ref{fig:projections}. The asymmetric LHCb kinematics 
causes that $x_1$ is rather large and $x_2$ is very small --
even smaller than 10$^{-5}$, much smaller than for other perturbative
partonic processes. 
We observe also a clear asymmetry in $q_{1T}$ and
$q_{2T}$. The $q_{2T}$ transverse momentum corresponding to small
$x_2$ is substantially larger than $q_{1T}$ of the large-$x_1$ gluon.
The low-$x$ gluon therefore transfers the bulk of the transverse momentum of the $\eta_c$ at large $p_T$.

We can exploit the good separation in $x_{1,2}$ to investigate the small-$x$ behaviour of the unintegrated glue. From our choice of UGDs, the 
parametrizations of Kutak are available only 
for $x < 0.01$, so we will use these UGDs only
for the small-$x$ gluon. To avoid a proliferation of plots, we will use the Kutak UGDs always together with the KMR UGD for the large-$x$ gluon.
A similar strategy was taken previously in Ref.~\cite{CS2018}.

The distributions for different UGDs in Fig.~\ref{fig:projections} are rather similar which makes are conclusions more universal.

%---------------------------------------------------------------------------
%----------------------------------------------------------------------------

\begin{figure}[!h]
    \centering
    \includegraphics[width=0.45\textwidth]{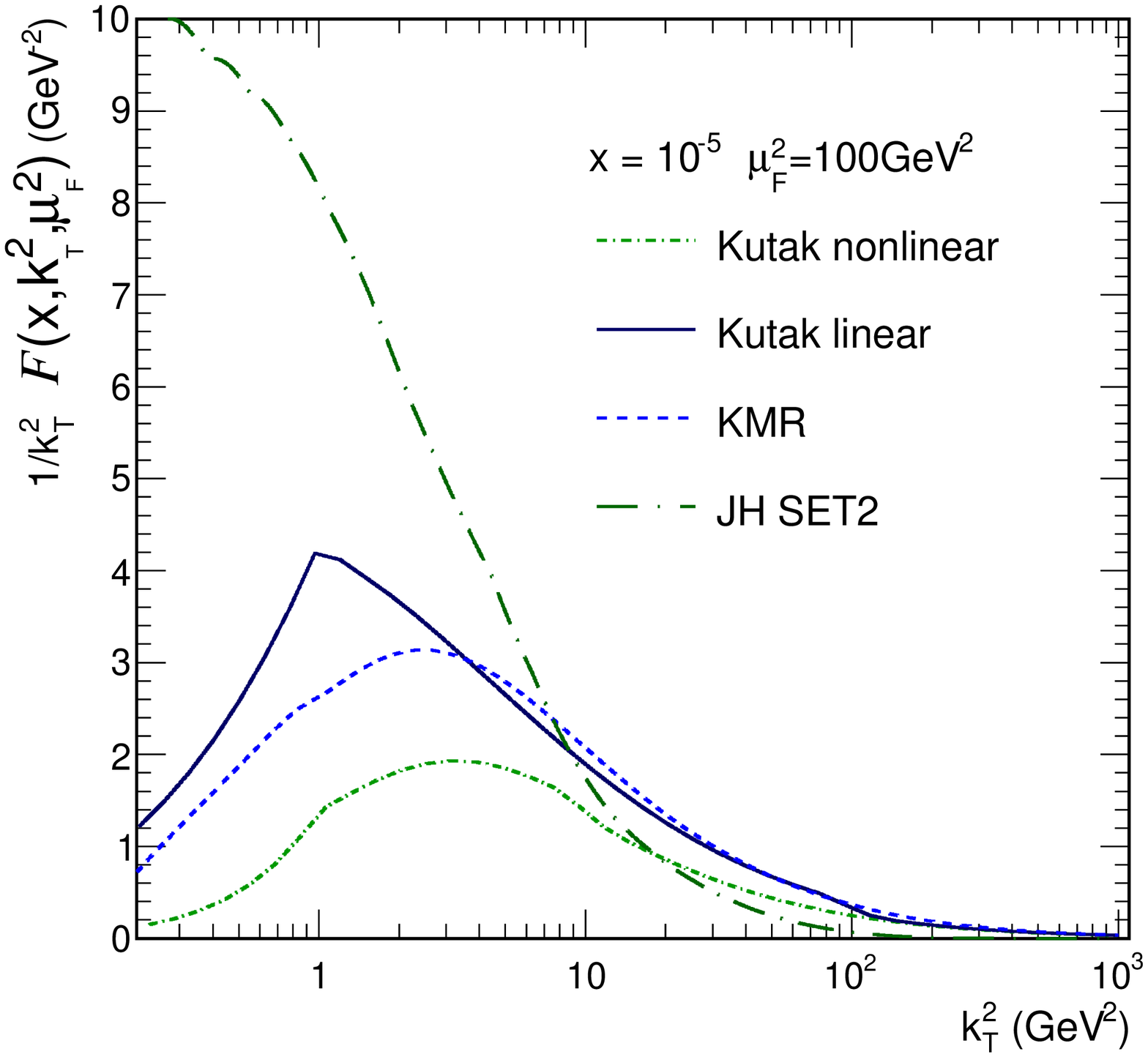}
    \includegraphics[width=0.45\textwidth]{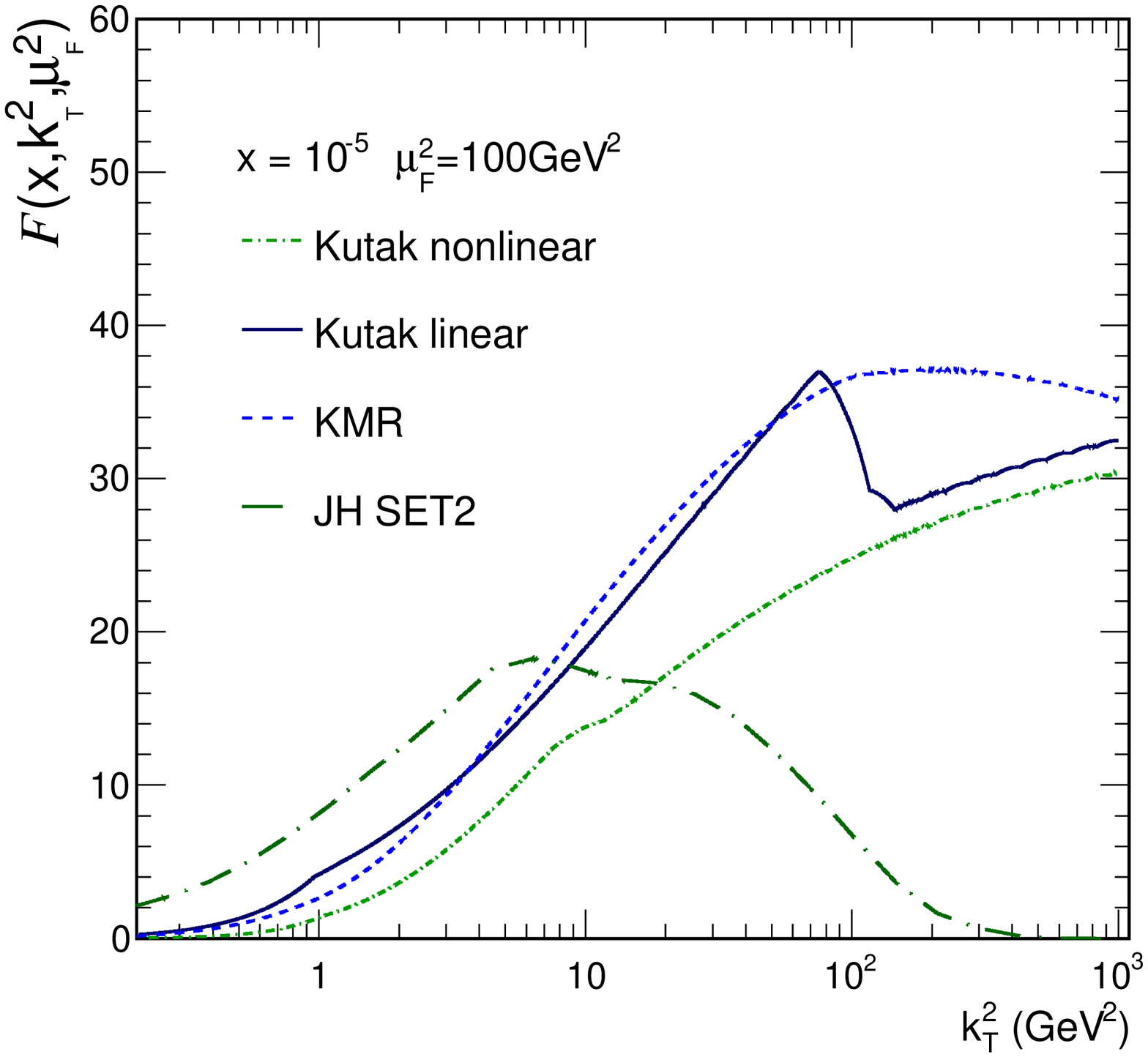}
    \caption{Unintegrated gluon densities for typical scale $\mu^2$= 100
      GeV$^2$ for $\eta_{c}(1S)$ production in proton-proton scattering 
at LHCb kinematics.}
    \label{fig:UGDs}
\end{figure}

Now we wish to show the behaviour of the different unintegrated
gluon distribution on gluon transverse momentum $k_T^2$ for small 
$x_2$ = 10$^{-5}$ and typical scale parameter $\mu^2 = 100 \, \rm{GeV}^2$
relevant for the LHCb experiment.
In the left panel of Fig.~\ref{fig:UGDs} we plot ${\cal{F}}(x,\bk^2,\mu_F^2)/\bk^2$, which corresponds to the distribution of the gluon transverse momentum squared $\bk^2$.
In the right panel the dimensionless UGD ${\cal{F}}(x,\bk^2,\mu_F^2)$ is plotted. 
We show all the UGDs used in the present work. The left panel of Fig.~\ref{fig:UGDs} better shows the behaviour at smaller $\bk^2 = k_T^2$, while the right panel emphasizes the large-$k_T$ tails.
We first observe, that the ``linear'' Kutak UGD looks quite
similar to the KMR UGD, although both are constructed by different procedures. They have in common, that by their construction, both procedures lead to integrated gluon distributions which well describe jet cross sections at the LHC.
The nonlinear Kutak UGD is considerably
smaller than the linear one, especially at low transverse momenta.
At very large transverse momenta the difference between linear and nonlinear UGDs becomes much smaller.
The Jung-Hautmann distribution does not have an extended tail in $k_T^2$ as the other distributions, it is however much larger at low gluon transverse momenta.
Can these different UGDs be tested in $\eta_c$ production?

%-----------------------------------------------------------------------------
%------------------------------------------------------------------------------

\begin{figure}[!h]
    \centering
    \includegraphics[width=0.45\textwidth]{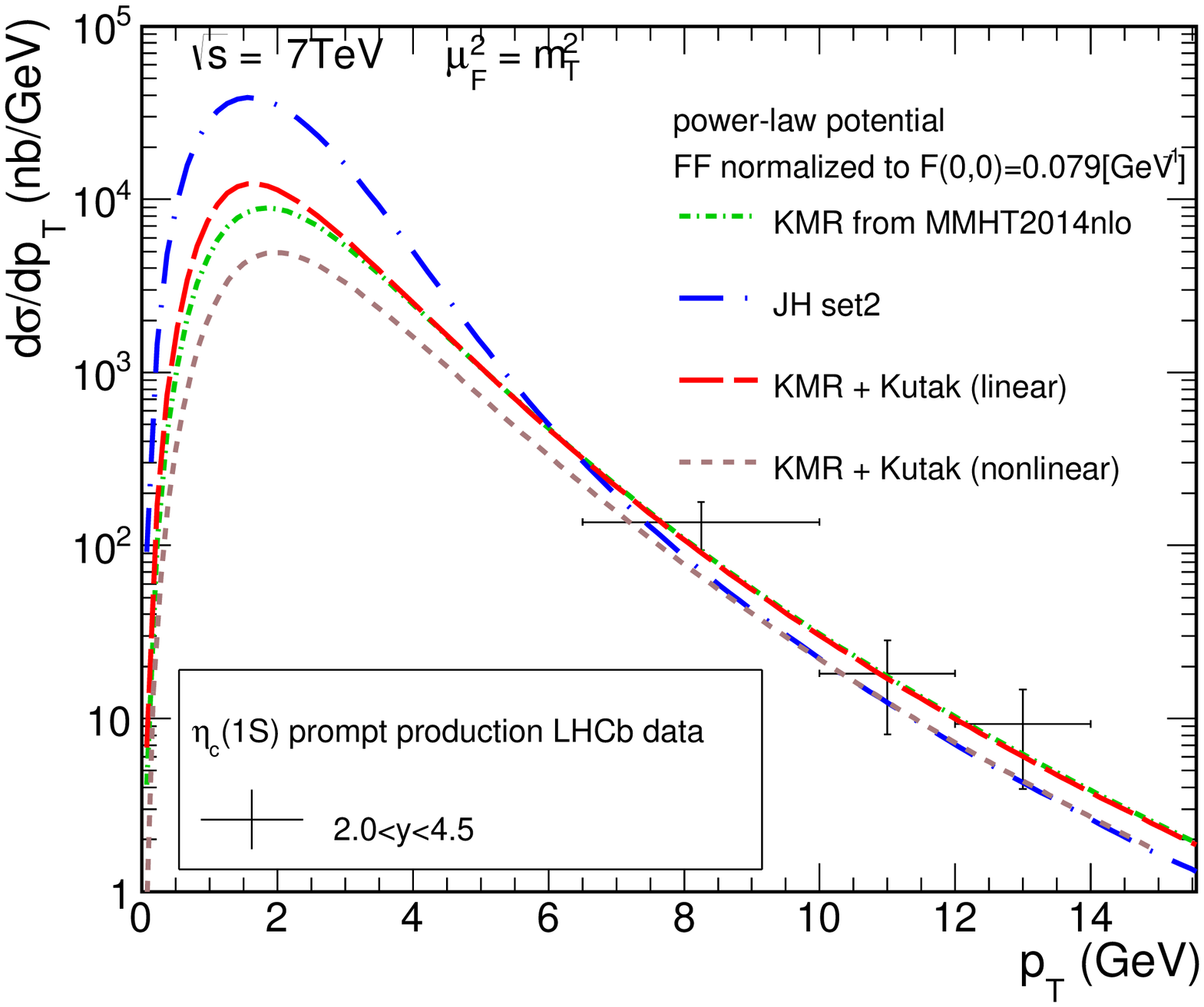}
    \includegraphics[width=0.45\textwidth]{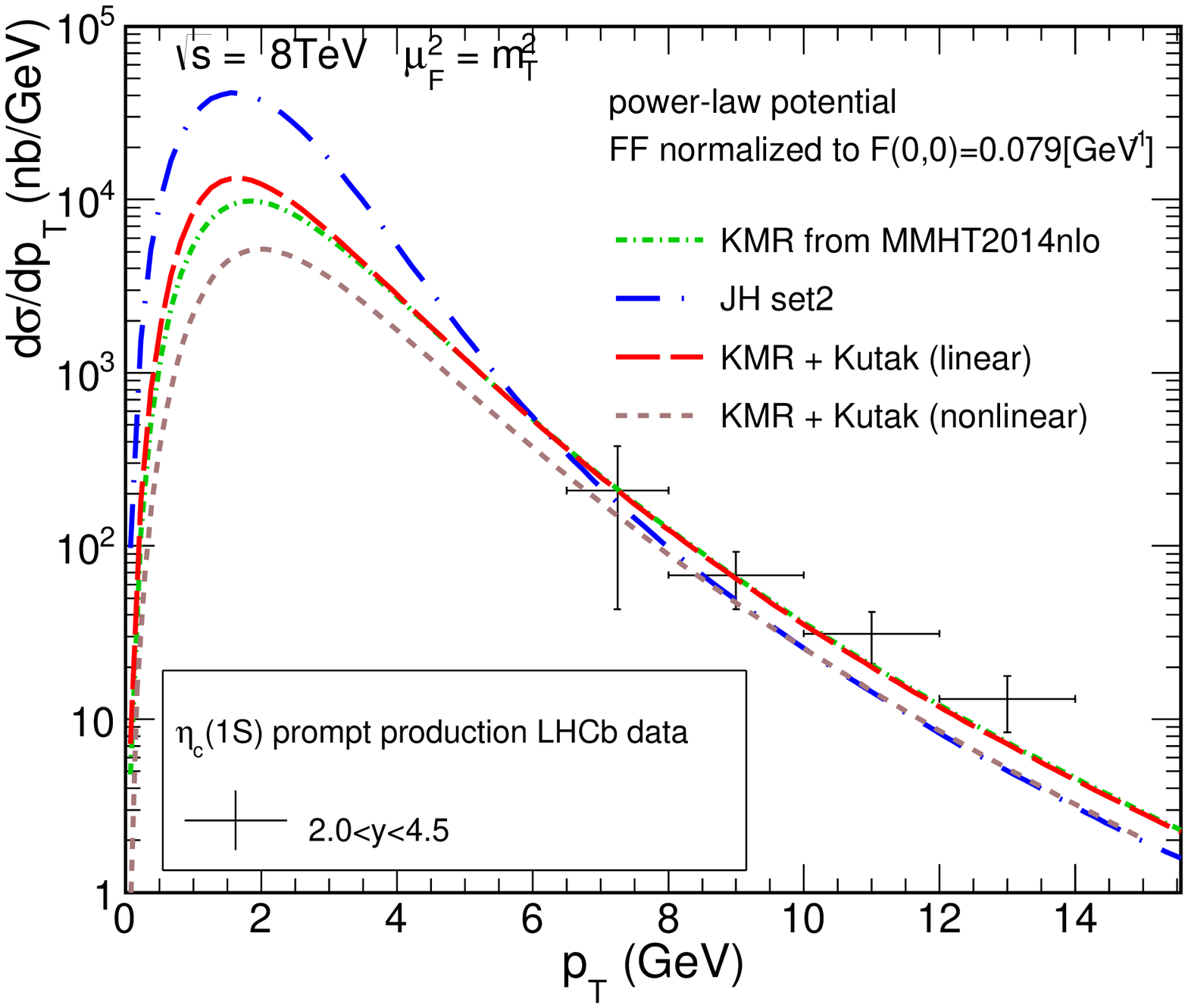}
    \includegraphics[width=0.45\textwidth]{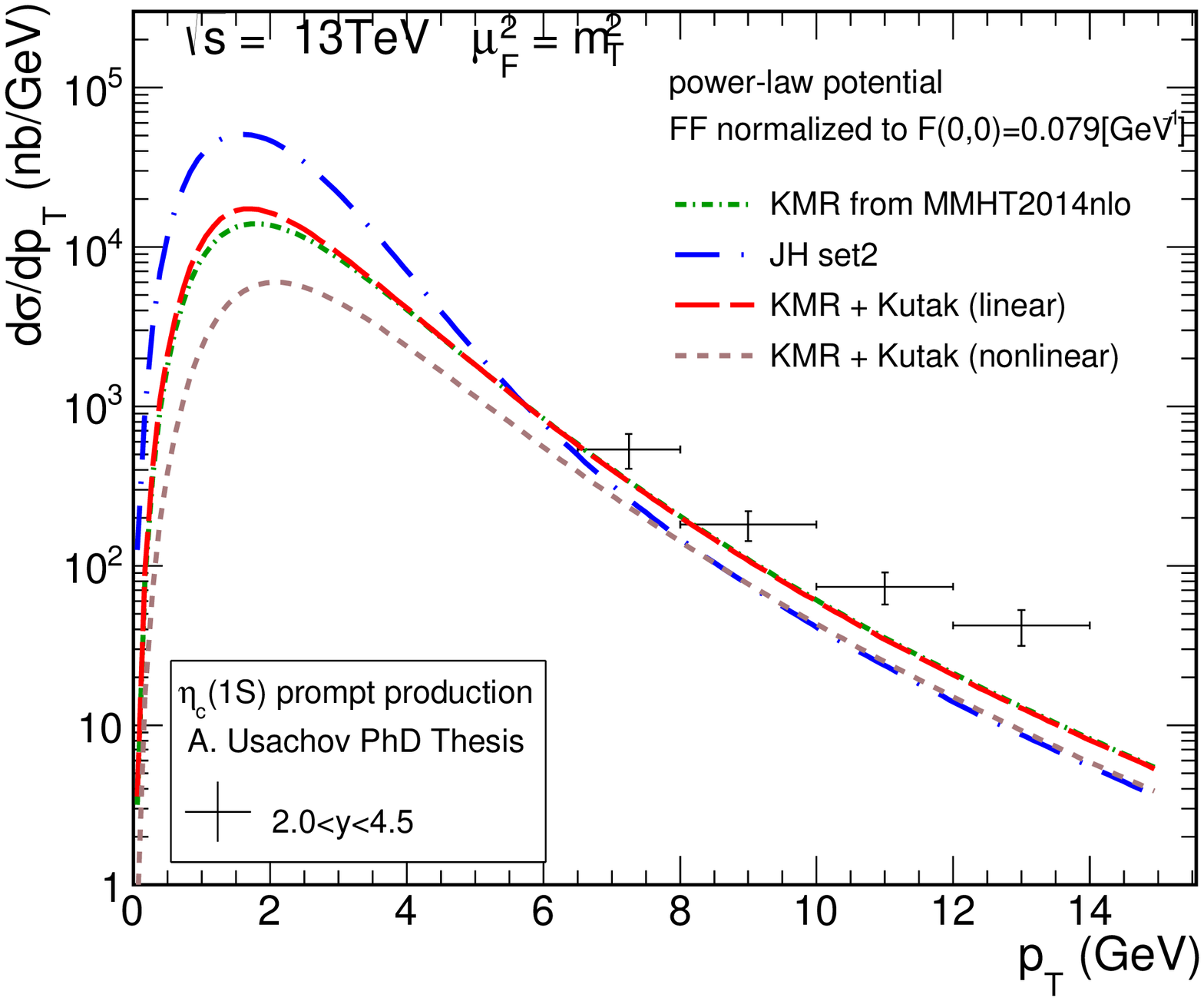}
    \caption{Differential cross section as a function of transverse momentum for prompt
      $\eta_{c}(1S)$ production compared with the LHCb data 
      \cite{Aaij:2014bga} for $\sqrt{s} =7, 8 \, \rm{TeV}$
      and preliminary experimental data \cite{Usachov:2019czc} 
     for $\sqrt{s}$ = 13 TeV.
     Different UGDs were used. Here we used the $g^* g^* \to \eta_c(1S)$ form factor calculated from the power-law potential.}
\label{fig:dsig_dpt_etac1S}    
\end{figure}

In Fig. \ref{fig:dsig_dpt_etac1S} we compare the results of the
$k_T$-factorization approach with the LHCb experimental data
for $\sqrt{s} = 7 \, \rm{and} \, 8 \, \rm{TeV}$ from Ref.~\cite{Aaij:2014bga} and with the data for $\sqrt{s} = 13 \, \rm{TeV}$ from the recent 
PhD thesis \cite{Usachov:2019czc}. The theoretical calculations use
an off-shell form factor normalized to the $\gamma \gamma$-decay width at NLO. The off-shell $g^* g^* \eta_c$ form factor was calculated using the LF wave function obtained for one of the potentials (the so-called power-law potential) in \cite{Cepila:2019skb}. It is up to the color factor proportional to the $\gamma^* \gamma^* \eta_c$ form factor obtained in \cite{Babiarz:2019sfa}. The dependence on the choice of the potential will be discussed below.

The description of data for  $\sqrt{s} = 7, 8 \, \rm{TeV}$ is 
reasonable for all UGDs. The theoretical results 
tend to be somewhat lower than the experimental data, especially at large $p_T$. 
The best description is obtained for the KMR UGD and the 
linear UGD by Kutak.

Our calculations fare a bit worse in the comparison to the data
at $\sqrt{s} = 13 \, \rm{TeV}$. Here all the UGDs give results
substantially below the data.
Please note, that we include only the direct production 
mechanism in the color-singlet channel.
A possible feed down from higher resonances, 
e.g. from the $h_c \to \eta_c \gamma$
radiative decay, is not included. See the recent 
Ref.~\cite{Baranov:2019joi} for an estimate which finds
a few percent contribution from the feed down.
We also do not include a possible color-octet 
contribution.

In general the difference between different UGDs is the largest at low $p_T$. Here one may e.g. expect effects related to
nonlinear evolution and gluon saturation. While it is true that the cross section peaks precisely in this most interesting region around $p_T \sim 2 \, \rm{GeV}$ it seems
to be exceedingly difficult to measure at $p_T < 6 \, \rm {GeV}$. At least this is true in the $p \bar p$ decay channel used by the LHCb collaboration.  Perhaps other decay channels
would be better in this respect. The $\gamma \gamma$ channel seems interesting as another option. A simulation of the signal
and background would be valuable in this context.

%----------------------------------------------------------------------------
%----------------------------------------------------------------------------

\begin{figure}[!h]
    \centering
    \includegraphics[width=0.45\textwidth]{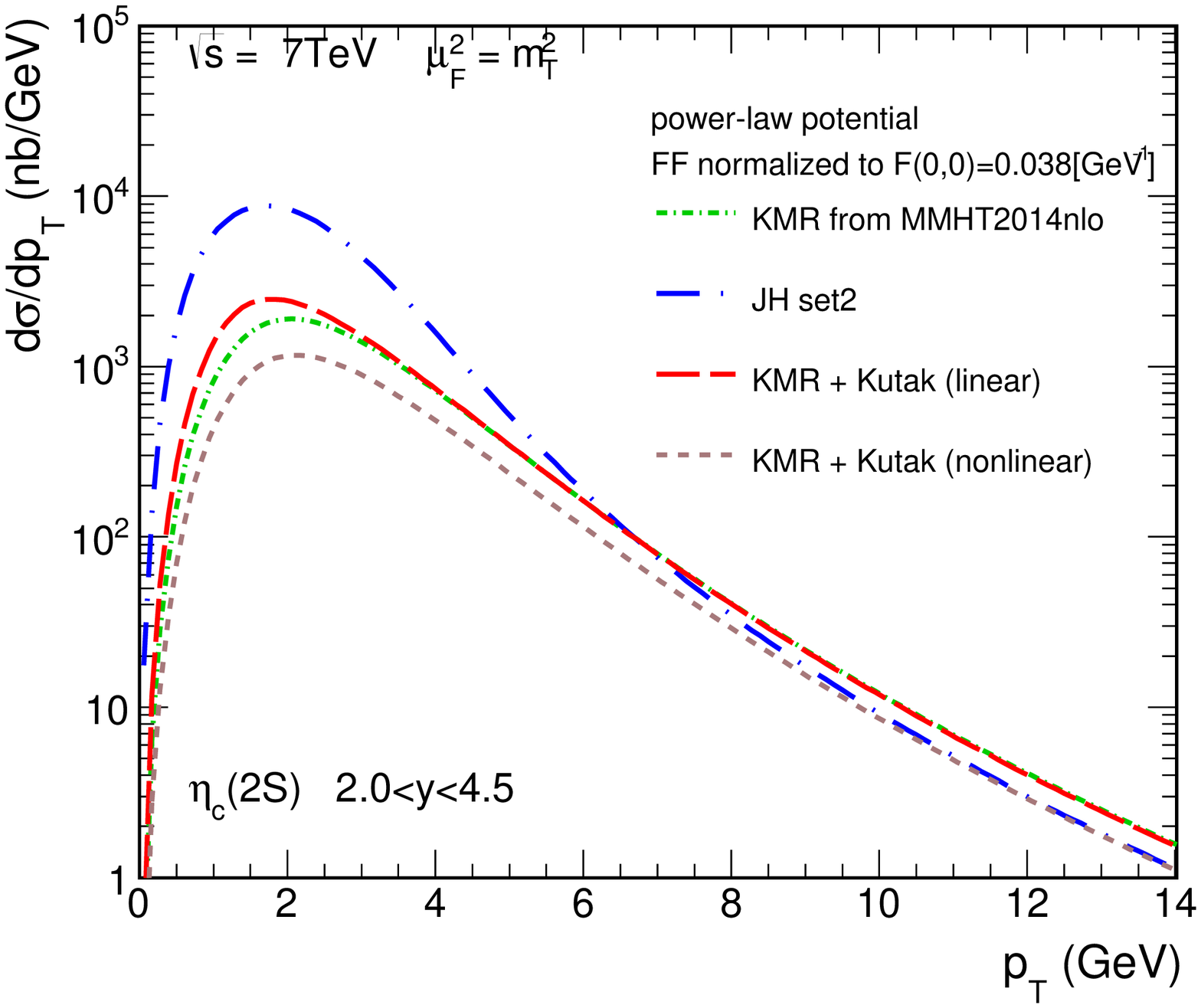}
    \includegraphics[width=0.45\textwidth]{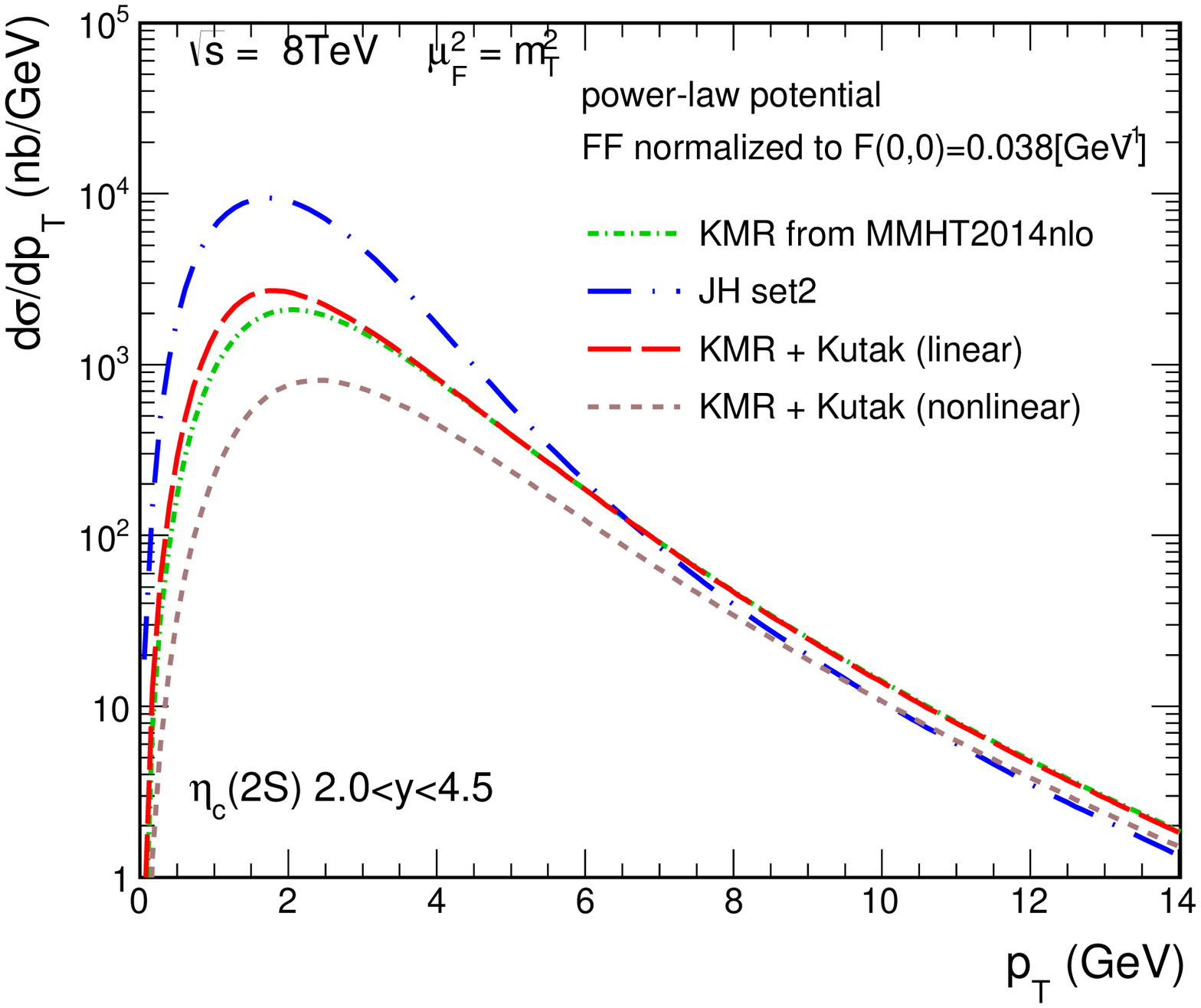}
    \includegraphics[width=0.45\textwidth]{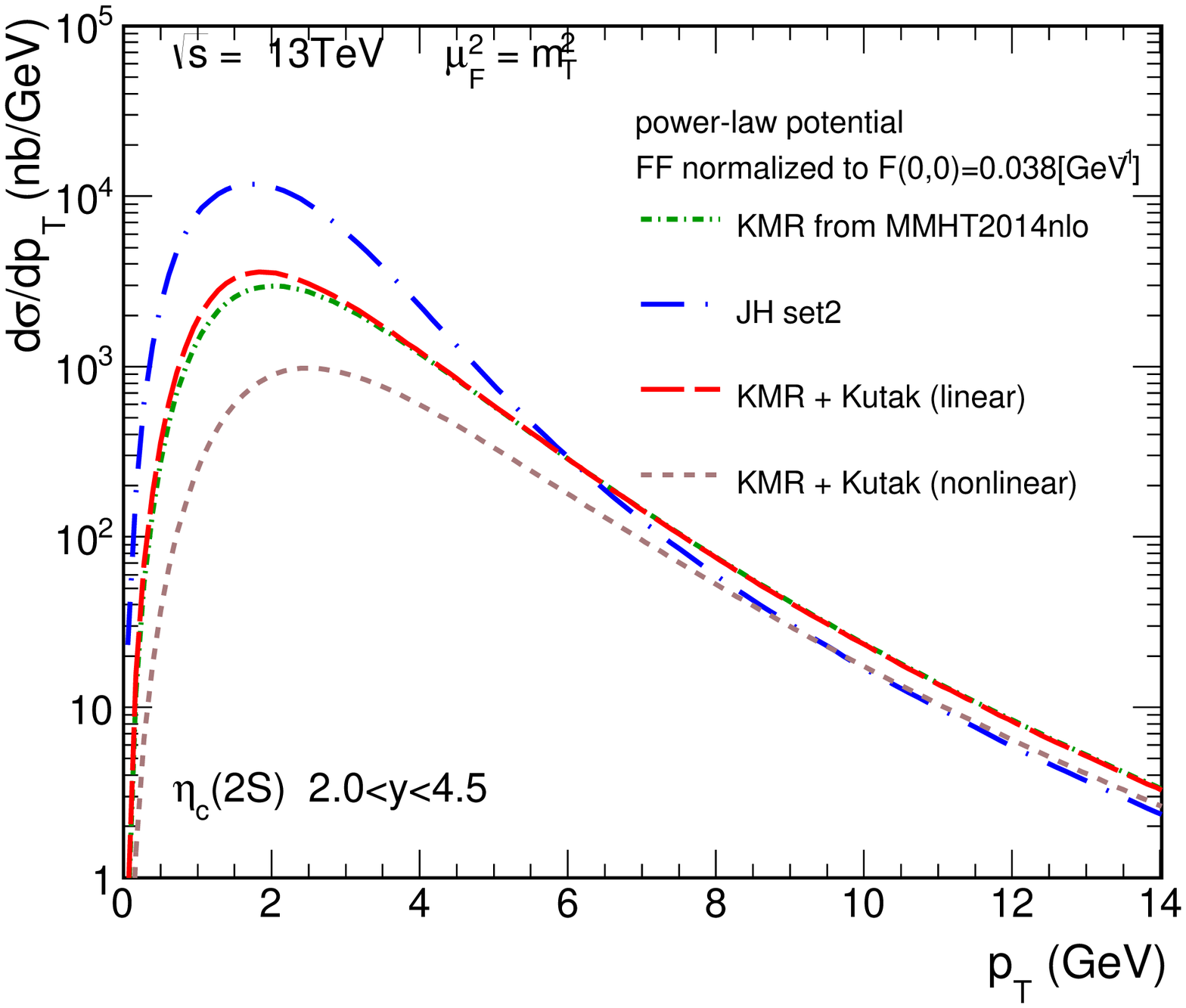}
    \caption{Differential cross section as a function of transverse
      momentum for prompt $\eta_{c}(2S)$
      production for $\sqrt{s} = 7, 8, 13 \, \rm{TeV}$.
}
    \label{fig:dsig_dpt_etac2S}
\end{figure}

In Fig.\ref{fig:dsig_dpt_etac2S} we show our predictions for the
so far unmeasured (at the LHC) $\eta_c(2S)$. In this calculation we also adjusted the value of $|F(0,0)|$ to the $\gamma \gamma$ width at NLO.
%experimental
%value of $F(q_{1T}^2 = 0, q_{2T}^2 = 0)$ $\gamma^* \gamma^* \to %\eta_c(2S)$ 
%transition form factor were used. 
The cross section is in a similar ballpark as for $\eta_c(1S)$ and the results for different UGDs
show a similar variation as for the $\eta_c(1S)$ production.

%----------------------------------------------------------------------------
%----------------------------------------------------------------------------

\begin{figure}[h!]
    \centering
    \includegraphics[width=0.45\textwidth]{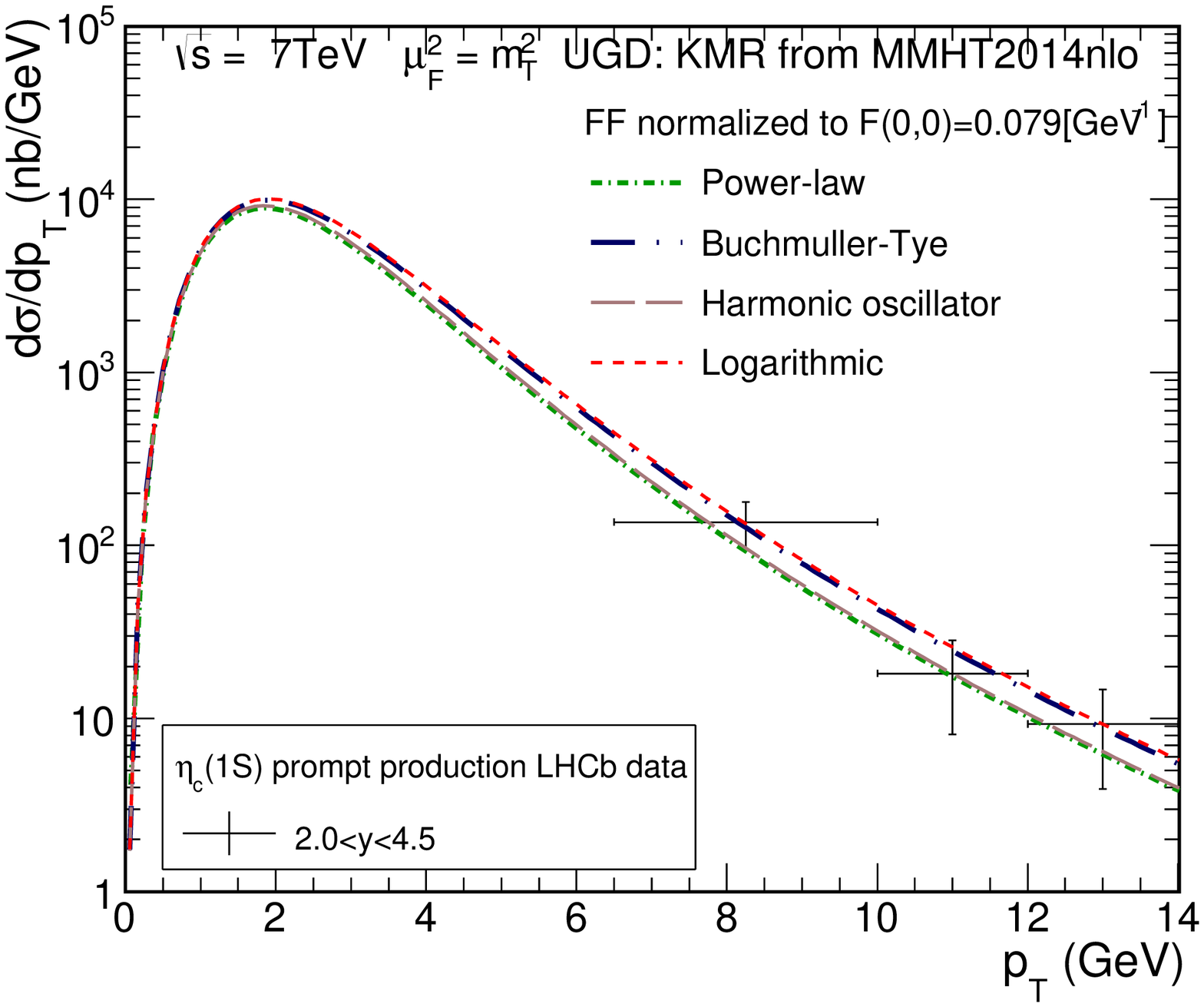}
    \includegraphics[width=0.45\textwidth]{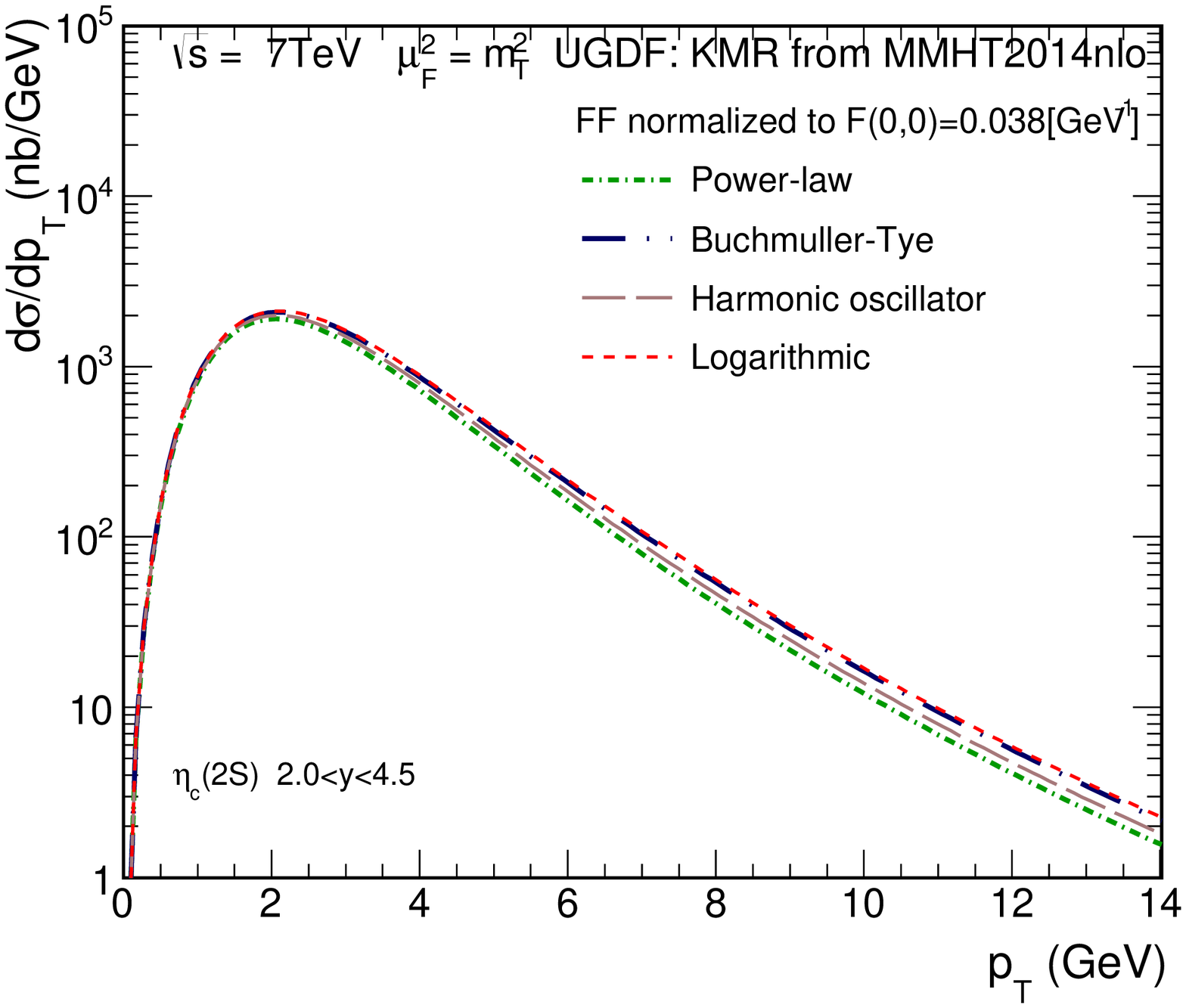}
    \caption{Transverse momentum distributions calculated with several 
     different form factors obtained from different potential models 
      of quarkonium wave function and
      one common normalization of $|F(0,0)|$ as explained in the text.}
    \label{fig:dsig_dpt_ff_exp}
\end{figure}

We now wish to turn to the dependence of our results on
the $c \bar c$ potential used for the calculation of the LF wave function.
In Ref.~\cite{Babiarz:2019sfa} the $\gamma^* \gamma^* \to \eta_c(1S,2S)$ transition form factors, which is closely related to the $g^* g^* \eta_c(1S,2S)$ form factors were obtained for different $c \bar c$ potentials. 
In Fig.\ref{fig:dsig_dpt_ff_exp} we show our results for different wave functions ($c \bar c$ potentials). 
To compare the influence the different wave functions have on the shape of the cross section, the respective form factors 
at the on-shell point, $|F(0,0)|$ were all adjusted to the same value dictated by the NLO expression for the experimental $\eta_c(1S,2S) \to \gamma \gamma$ decay width.

%----------------------------------------------------------------------------
%-----------------------------------------------------------------------------

\begin{figure}[!h]
    \centering
    \includegraphics[width=0.45\textwidth]{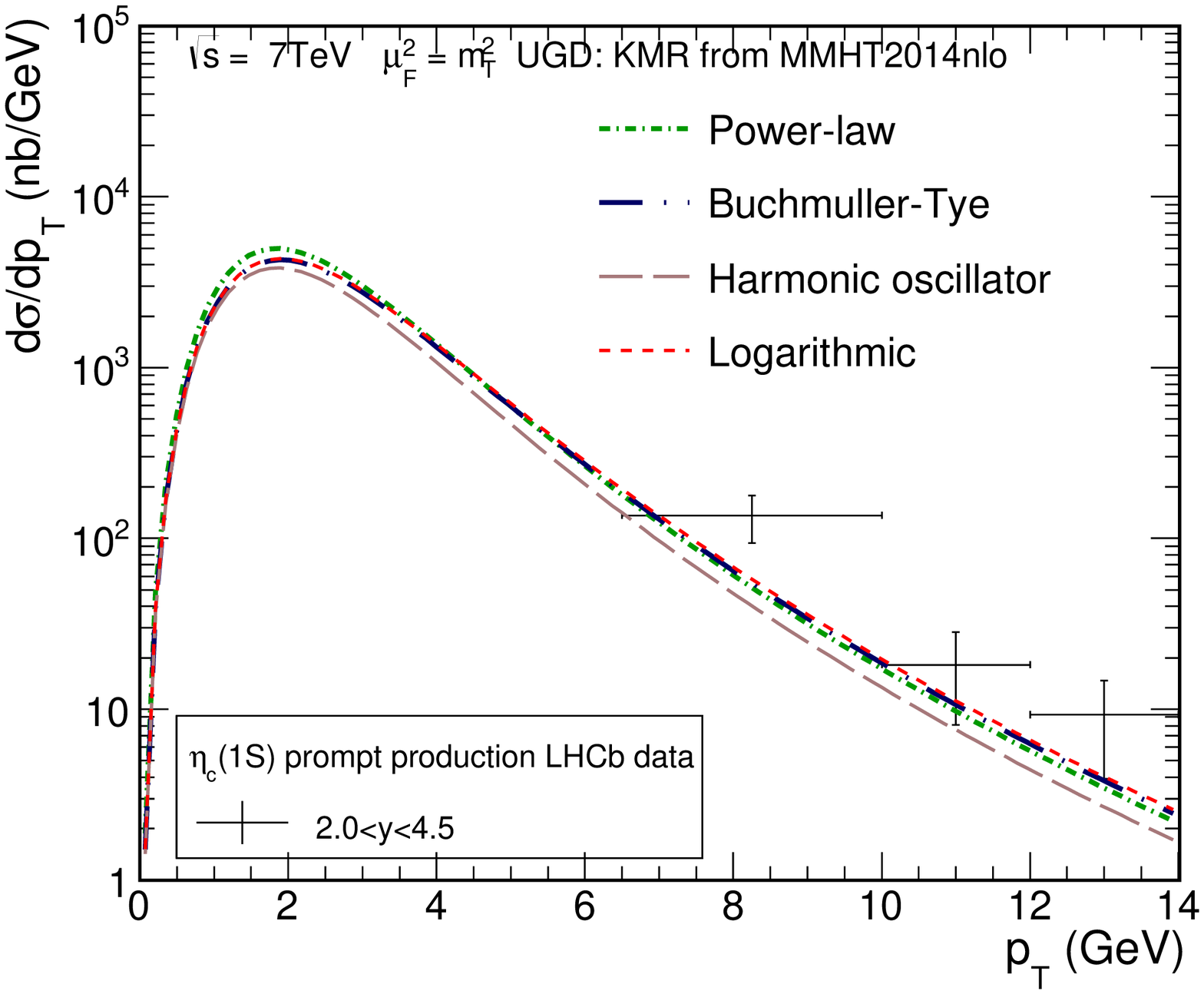}
    \includegraphics[width=0.45\textwidth]{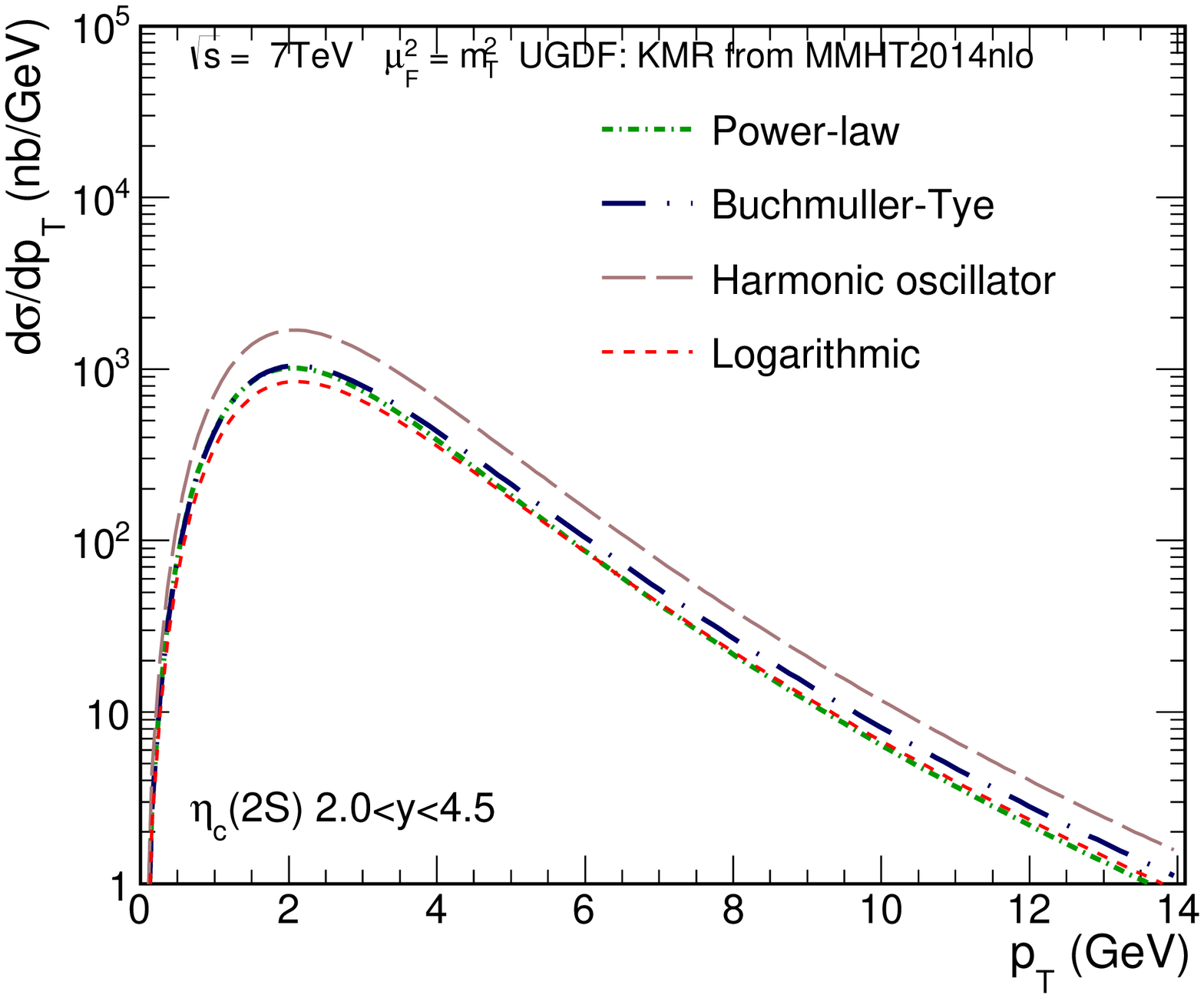}
    \caption{Distributions calculated with several different form factors obtained from different potential models of quarkonium.}
    \label{fig:dsig_dpt_ff}
\end{figure}

In Fig.\ref{fig:dsig_dpt_ff} we relax the normalization to the values predicted by the different potentials. 
Here the spread of results is bigger than in the previous case.
We need to caution the reader, that generally the results 
from the phenomenological potentials undershoot the experimental widths.
However, we wish to notice that experimental decay widths are known only with some precision \cite{PDG}. In \cite{Metreveli:2007sj} a different values was measured.
It appears, that the behaviour of the off-shell form factors 
at large $\bq_i^2$ is a more reliable result than their value at the on-shell point, though.

%----------------------------------------------------------------------------
%-----------------------------------------------------------------------------

\begin{figure}[h!]
    \centering
    \includegraphics[width=0.45\textwidth]{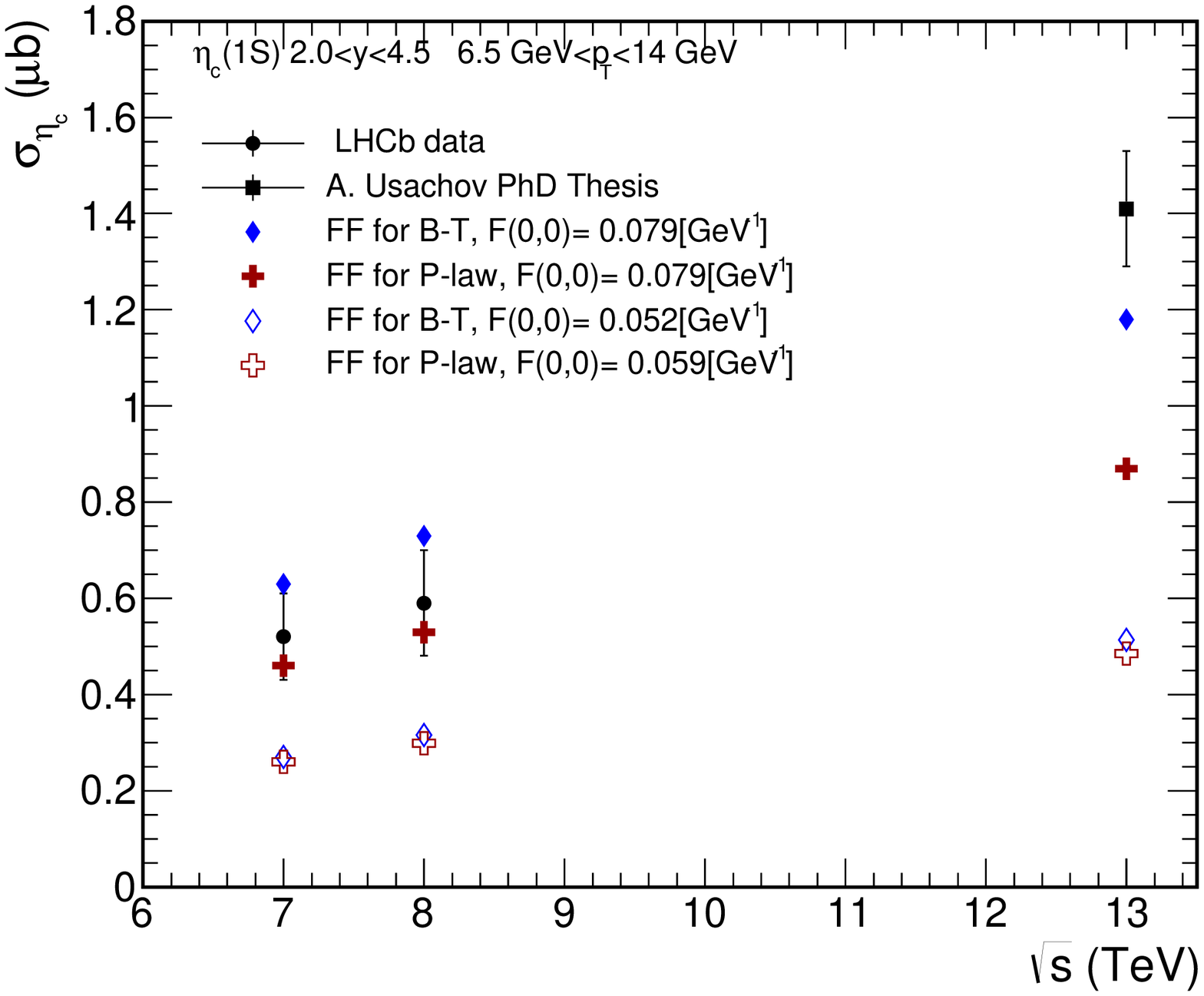}
    \caption{The integrated cross section computed within LHCb range of $p_{T}$ and $y$ with our transition form factors, compared to experimental values. Here red crosses represent values for Buchm\"uller-Tye potential (B-T) and deltoids for Power-law potential (P-law).  }
    \label{fig:sig_tot_1S}
\end{figure}

In Fig.\ref{fig:sig_tot_1S} we show the integrated cross section within the LHCb cuts for the three different energies
measured up to now. In this plot we used the KMR UGD.
We observe that the data appear to indicate a faster than linear rise of the cross section, while the theoretical calculations
predict a slightly slower rise of the cross section with $\sqrt{s}$.
We checked that similar results are obtained with other UGDs.
Once more we remind of the missing feed-down component.
A possible color-octet component is not expected to change
the energy dependence, as it is driven by the similar $g^* g^* \to c \bar c$ process.

%-----------------------------------------------------------------------------
%-----------------------------------------------------------------------------

\begin{figure}[h!]
    \centering
    \includegraphics[width=0.45\textwidth]{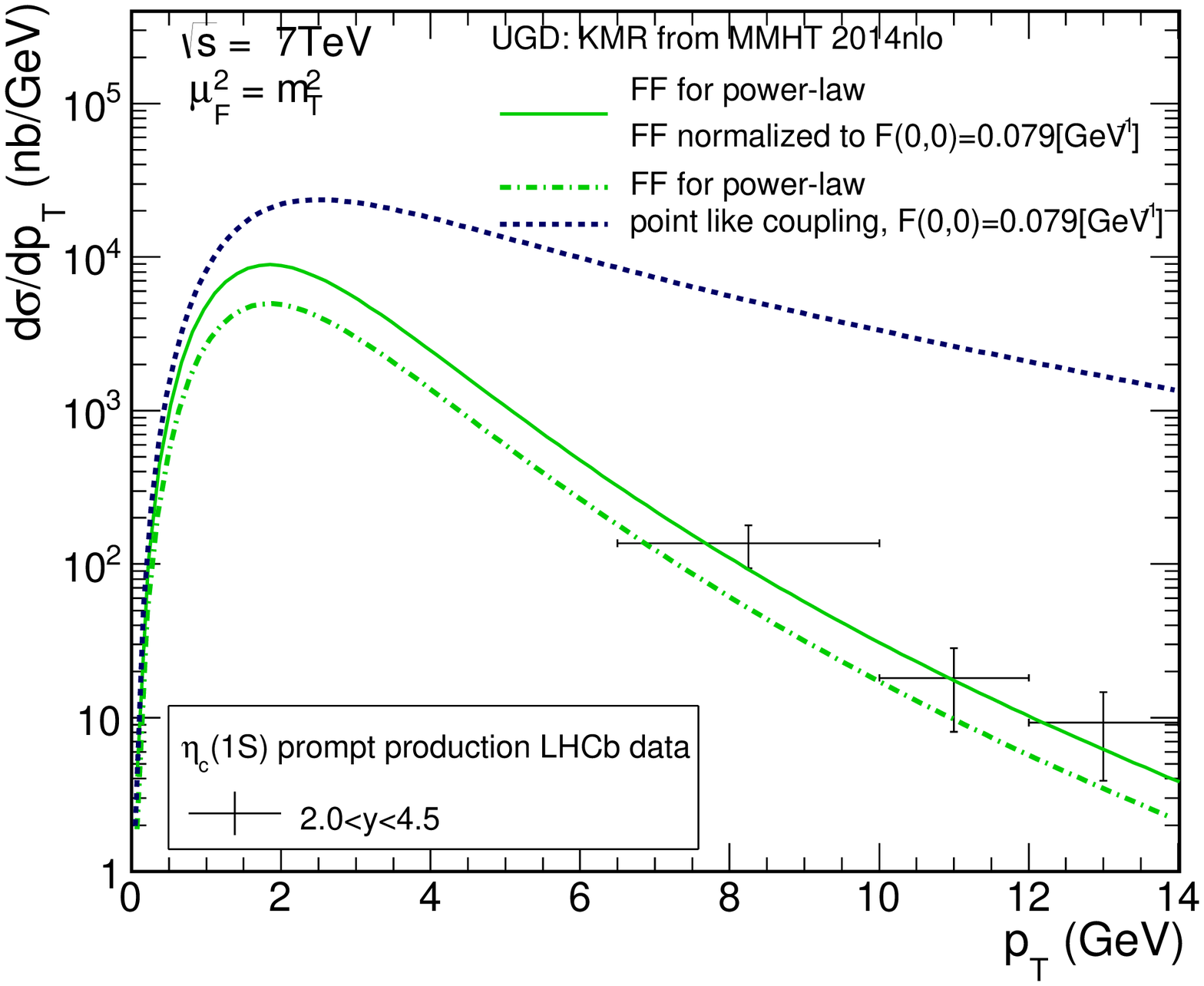}
    \caption{Comparison of results for two different transition form
      factor, computed with the KMR unintegrated gluon distribution.
      We also show result when the $(q_{1T}^2,q_{2T}^2)$ dependence
      of the transition form factor is neglected.}
    \label{fig:Different_coupling}
\end{figure}

It is interesting to investigate what is the role of the off-shell form factor. For example in the approach of
Ref.~\cite{Boer:2012bt} gluon virtualities are neglected
in the hard matrix element.
The curves in Fig.~\ref{fig:Different_coupling} clearly show that the effect of the inclusion of gluon virtualities
in the transition form factor is essential and cannot be neglected.

%------------------------------------------------------------------------------
\begin{figure}[h!]
    \centering
    \includegraphics[width=0.45\textwidth]{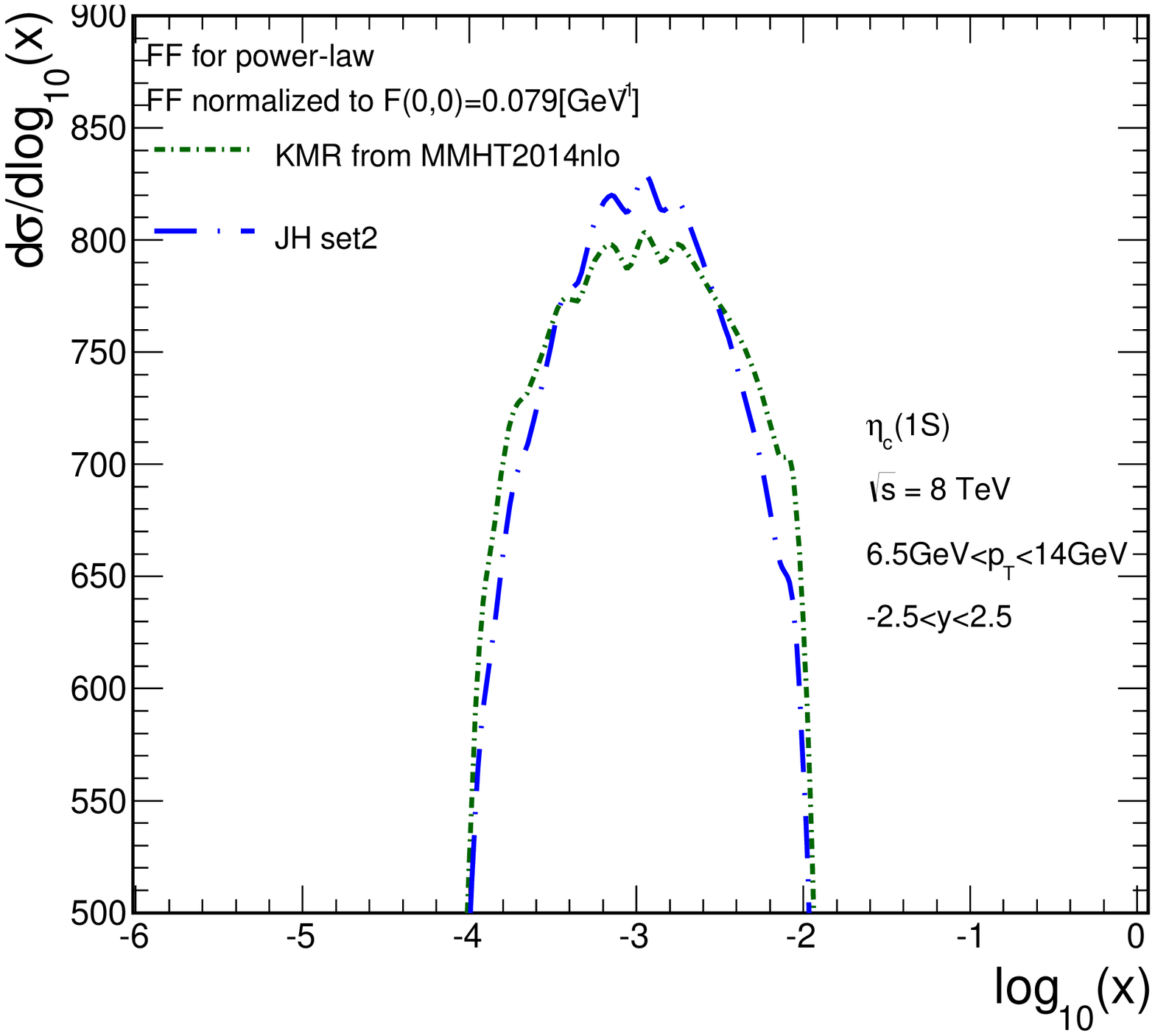}
    \includegraphics[width=0.45\textwidth]{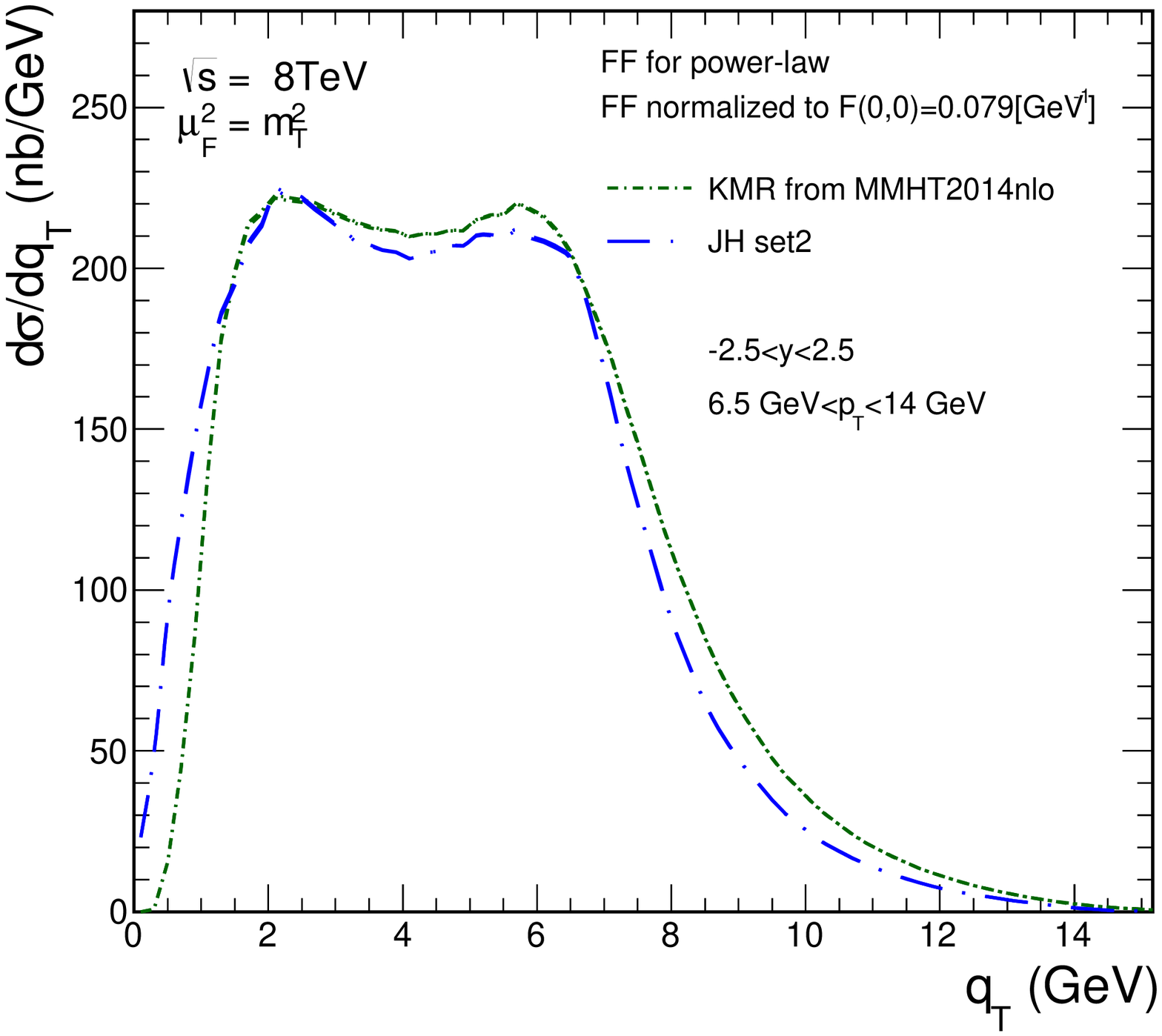}
    \caption{Distribution in $\log_{10}(x_1)$ or $\log_{10}(x_2)$ (left
      panel) and distribution in $q_{1T}$ or $q_{2T}$ (right panel) for
      ATLAS or CMS conditions.}
    \label{fig:kinematics_ATLAS}
\end{figure}

%----------------------------------------------------------------------------
Up to now we concentrated on the kinematics of the LHCb experiment. Could $\eta_c$ quarkonia be measured by other experiments at the LHC?
In Fig.\ref{fig:kinematics_ATLAS} we show the ranges of $x_i$ and $q_{Ti}$ carried by gluons for a rapidity interval
$-2.5 < y < 2.5$ typical of the central detectors of ATLAS
or CMS.
As in the case of LHCb, we assumed a lower cut on transverse momentum,
$p_T > 6.5 \, \rm{GeV}$. 
In the center-of-mass rapidity interval symmetric around zero,
of course both UGDs enter symmetrically and therefore the
distributions of $x_1$ and $x_2$ coincide, as do the 
ones for $q_{1T}$ and $q_{2T}$
We find that in the central rapidity region one would test
$x_1, x_2 \sim$ 10$^{-4}$ - 10$^{-2}$ i.e. the region where the gluon is already known reasonably well from the HERA experiments. 
The $q_{iT}$-distribution has a large plateau at perturbatively large values, so that we suppose that the predictions of
the KMR UGD should be reliable in this case.

%----------------------------------------------------------------------------
%----------------------------------------------------------------------------

\begin{figure}[h!]
    \centering
    \includegraphics[width=0.45\textwidth]{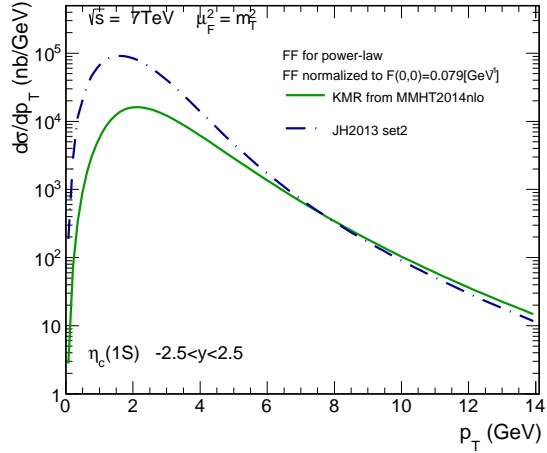}
    \caption{Transverse momentum distribution of prompt $\eta_{c}$(1S)
      for $-2.5<y<2.5$ and $\sqrt{s}$ =7 TeV.}
    \label{fig:dsig_dpt_ATLAS}
\end{figure}

The corresponding transverse momentum distributions are shown 
in Fig.\ref{fig:dsig_dpt_ATLAS} for two different UGDs reliable for
this region of longitudinal momentum fractions and gluon transverse
momenta. Notice that because there is some contribution of $x > 0.01$, we cannot use here the Kutak UGDs, which are unavailable
at these $x$-values.

%----------------------------------------------------------------------------
%----------------------------------------------------------------------------

%For completeness we wish to show also situation at midrapidities relevant for ATLAS,CMS experiments. We show corresponding $x_{1}, x_{2}$ (left panel and $q_{1T}$, $q_{2T}$ (right panel) distributions. For the same cut on transverse momentum of $\eta_{c}$ as for LHCb we get $x_{1}$, $x_{2} \sim 10^{-3}$ and typical $q_{1T}, q_{2T} \sim 4 GeV$.

%--------------------------
\section{Conclusions}
\label{Sect:conclusions}
%--------------------------

In the present paper we have discussed in detail the production of pseudoscalar  $\eta_c$ quarkonia in proton-proton collisions at the LHC energies. 
The calculations have been performed in the $k_T$-factorization approach using $g^*  g^* \to \eta_c$ vertex calculated from the $c \bar c$ wave functions. 
The latter vertices are closely related to the 
$\gamma^* \gamma^* \to \eta_c$ transition form factors which were
obtained recently in \cite{Babiarz:2019sfa} and which we used
in the present work.

We have used different unintegrated gluon distributions available in the literature:
firstly the  KMR UGD, which effectively includes higher-order effects of the collinear approach, secondly
two UGDs by Kutak which were obtained by using linear and nonlinear small-$x$ evolution equations, and thirdly the Jung-Hautmann UGD from 2013, obtained from a fit to HERA data in a CCFM approach. 

We have calculated transverse momentum distributions of both 
$\eta_c(1S)$ and $\eta_c(2S)$ charmonia.
The results of the $\eta_c(1S)$ have been compared with the experimental data obtained by the LHCb collaboration for $\sqrt{s}  = 7, 8 \, \rm{GeV}$ \cite{Aaij:2014bga} and to a measurement at
$\sqrt{s} = 13 \, \rm{TeV}$ published in 
a PhD thesis \cite{Usachov:2019czc}.

A quite good agreement with the data was obtained with the KMR UGD.
% Kutak UGD is valid only for small values of longitudinal momentum
%fraction so cannot be used for both $x_1$ and $x_2$. 
%We performed calculations using the KMR UGD
%for the large-$x$ gluon and the Kutak UGDs for the low-$x$ gluon. Here the linear Kutak distribution also gives a reasonable description of the data.
We have shown the range of $x_1, x_2$ and gluon transverse momenta $q_{1T}, q_{2T}$ 
probed in the kinematics of the LHCb experiment \cite{Aaij:2014bga,Usachov:2019czc}.
For the LHCb experiment  one of the $x$-values, $x_1 \in (10^{-2}, 10^{-1})$ is large, while the second one, $x_2 \in (10^{-5}, 10^{-4})$ takes very small values. 
For the LHCb experiment we have shown also a large asymmetry in  $q_{1T}$ (larger) and $q_{2T}$ (smaller).
It turns out that at large $p_T$ of the meson the bulk of transverse momenta is transferred by the small-$x$
gluon.
The Kutak UGD cannot be used for the range of $x_1$ relevant for the LHCb experiment. Therefore in this case we have used two different UGDs:
Kutak UGD for small $x_2$ and KMR  UGD for large $x_1$.  
We have used both the linear and nonlinear UGDs
of Kutak.  The mixed UGD scenario with the linear Kutak  UGD leads to very similar transverse momentum distribution of $\eta_c$ as that for using the KMR UGD on both sides. 
Indeed in the considered range of $x$ and $k_T$ the linear Kutak UGD is very similar as the KMR UGD. 
The nonlinear version of the Kutak UGD leads to smaller cross sections, especially for small $\eta_c$ transverse momenta.

A measurement of $\eta_c$ at low transverse momenta in the LHCb kinematics would therefore be very valuable in the context of searching for nonlinear effects and onset of gluon saturation. One could, for example, consider another measurement for the $\gamma \gamma$ decay channel.

We have shown that it is crucial to include the dependence
on gluon virtualities in the $g^* g^* \to \eta_c(1S,2S)$ 
vertex. We have also discussed uncertainties related to the $g^* g^* \to \eta_c$(1S,2S) form factor.
We have shown results of calculations with the form factor obtained from different  $c \bar c$ potentials from the literature. The associated uncertainty is somewhat smaller than that related  to the choice of UGD.
A better measurement of $\eta_c$(1S,2S) $\to \gamma \gamma$ would be important in this context.

%At low transverse momenta of $\eta_c$ there would be also a sizeable uncertainty due to 
%the prescription of $\alpha_s$ for low $q_{1T}$ or/and $q_{2T}$. 
%For the LHCb experiment this is much less serious problem. 

%-------------------------------------------------------------------------
\section*{Acknowledgements}
We would like to thank Krzysztof Kutak for a discussion on UGDs and for providing the numerical grids of his UGDs.
This study was partially supported by the Polish National
    Science Center grant UMO-2018/31/B/ST2/03537 and by the Center for
Innovation and Transfer of Natural Sciences and Engineering Knowledge in Rzesz\'ow.

%-------------------------------------------------------------------------

\end{document}